\documentclass[%
reprint,
amsmath,amssymb,
aps,
prb,
floatfix,
]{revtex4-2}
\usepackage{graphicx,xcolor}
\definecolor{darkblue}{RGB}{0,0,150}
\definecolor{nightblue}{RGB}{0,0,100}
\newcommand{\markup}[1]{\textcolor{black}{#1}}
\usepackage{mathrsfs,dsfont,mathtools}

\usepackage{dcolumn}% Align table columns on decimal point
\usepackage{bm}% bold math
\usepackage[
colorlinks,
citecolor=darkblue,
linkcolor=darkblue,
urlcolor=nightblue]{hyperref}% add hypertext capabilities
\usepackage[english]{babel}
\usepackage[babel,kerning=true,spacing=true]{microtype}

\usepackage{feynmp-auto}

\renewcommand{\Re}[1]{\textrm{Re}\left( #1 \right)}
\renewcommand{\Im}[1]{\textrm{Im}\left( #1 \right)}
\newcommand{\bk}{\mathbf{k}}

\newcommand{\tlo}{\tilde{\Omega}}
\newcommand{\lra}{\leftrightarrow}
\newcommand{\nd}{2\textsuperscript{nd}}
\newcommand{\vare}[2]{\varepsilon_{#1 #2}}
\newcommand{\dO}{\delta\omega}
\definecolor{DarkRed}{RGB}{100,0,0}
\AtBeginDocument{\renewcommand{\natexlab}[1]{#1}}% <--- the fix

\begin{document}
	
	\title{Unifying semiclassics and quantum perturbation theory at nonlinear order}
	
	\author{Daniel Kaplan}
	\email{daniel.kaplan@weizmann.ac.il}
	\author{Tobias Holder}
	%\email{tobias.holder@weizmann.ac.il}
	\author{Binghai Yan}
	\affiliation{Department of Condensed Matter Physics,
		Weizmann Institute of Science,
		Rehovot 7610001, Israel}
	
	\date{\today}
	
	\begin{abstract}
		Nonlinear electrical response permits a unique window into effects of band structure geometry. It can be calculated either starting from a Boltzmann approach for small frequencies, or using Kubo's formula for resonances at finite frequency. However, a precise connection between both approaches has not been established. Focusing on the second order nonlinear response, here we show how the semiclassical limit can \markup{be} recovered from perturbation theory in the velocity gauge, provided that finite quasiparticle lifetimes are taken into account. We find that matrix elements related to the band geometry combine in this limit to produce the semiclassical nonlinear conductivity. 
		We demonstrate the power of the new formalism by deriving a quantum contribution to the nonlinear conductivity which is of order $\tau^{-1}$ in the relaxation time $\tau$, which is principally inaccessible within the Boltzmann approach. We outline which steps can be generalized to higher orders in the applied perturbation, and comment about potential experimental signatures of our results. 
	\end{abstract}
	
	\maketitle
	
	\section{Introduction}
	It is well-known that nonlinear electrical response~\cite{Belinicher1980,vonBaltz1981,Boyd2003,Kraut1979,vonBaltz1981,Sipe2000} allows \markup{one}
	to probe properties of quantum materials 
	%-- through the interaction of light and matter -- %
	which are fundamentally inaccessible in the linear regime \cite{Bhalla2020,Ahn2020,Holder2020,Kaplan2020,Watanbe2020,Ahn2021,Kaplan2021a,Watanbe2021,chaudhary2022}. 
	%% Add one or more sentences about electrical conductivity %%
	However, the theoretical treatment of nonlinear electrical response has so far been divided into two seemingly unrelated approaches: semiclassical~\cite{Sundaram1999,Culcer2005,Chang2008,Bhalla2020,Bhalla2021,Atencia2022,Sturman1992,Sturman2019,Fradkin1986}, or quantum perturbative~\cite{Ward1965,Kraut1979,vonBaltz1981, Boyd2003,Parker2019,Holder2020,Watanbe2021b,Watanbe2022}. 
	For a microscopic understanding of the light-matter interaction it is imperative to understand the relation between these two approaches and how they can be related to each other mathematically. Somewhat unexpectedly, this basic question has not been resolved in the literature, despite receiving increased attention recently~\cite{Nastos2006,Cheng2015,Ventura2017,Passos2018,Michishita2021}.
	
	The semiclassical approach starts by considering an equilibrium distribution of electrons in a Bloch band, and then time evolving this distribution function according to a Boltzmann equation. In this approach, momentum-relaxing processes like scattering by impurities or phonons are introduced through a collision term~\cite{Niu1996,Gao2019, Niu2019, Du2021}, which is often approximated in its entirety by an averaged relaxation time $\tau$. 
	%This relaxation time can be calculated from models of scatterers or impurities, but is often taken to be phenomenological. 
	The validity of the semiclassical approach is considered to be limited to metallic systems and frequencies $\omega \ll E_g$, where $E_g$ is the single-particle gap which separates the ground state from excited states \cite{Deyo2009,Sturman1992,Sundaram1999}.
	In contrast, the quantum perturbative picture is assumed to be relevant whenever the frequency of the incoming light is large ($\omega > E_g$)~\cite{Boyd2003}. The interaction of the perturbing field with matter is implemented via one of two gauge choices: the length gauge \cite{Aversa1995,Sipe2000}, in which the Hamiltonian is transformed as $H \to H + e \mathbf{E}(t) \mathbf{\hat{r}}$, where $\mathbf{E}(t)$ is the applied electric field, and $\mathbf{r}$ is the position operator; or in the velocity gauge \cite{Parker2019} where the crystal momentum is shifted through minimal coupling, and the Hamiltonian accordingly becomes $H(\bk) \to H(\bk + e \mathbf{A})$, where $\mathbf{A}$ is the (time-dependent) vector potential. In the quantum perturbative picture, lifetimes can be added in the form of a finite fermionic self-energy in order to truncate divergences and broaden the resonance factors~\cite{Boyd2003}. However, in the literature this lifetime is often removed in the final steps of the calculation by taking $\tau \to \infty$ \cite{Kraut1979,Boyd2003,Sipe2000,LGao2021,Lihm2022}, thereby eliminating its effect.
	
	Several issues continue to plague the quantum perturbative approach: some formulations of the velocity gauge formalism have been shown to exhibit a spurious zero-frequency divergence \cite{Ventura2017,Passos2018, Mikhailov2014,Mikhailov2016}, which has to be eliminated by imposing sum rules or symmetry restrictions. In the length gauge, multiple prescriptions have been proposed for the implementation of finite lifetimes, and to date it remains unclear which one is most appropriate~\cite{Cheng2014,Cheng2015,Cheng2017}. Additionally, the resulting expressions after taking the limit $\tau \to \infty$ strongly depend on time reversal symmetry (TRS) \cite{Aversa1995, Nastos2006}. TRS has also been invoked in some studies~\cite{Aversa1994,Aversa1995,Cheng2015,Mikhailov2016} to justify the elimination of explicitly lifetime dependent terms. 
	We have recently proposed a physical interpretation for the lifetimes introduced in velocity gauge expressions~\cite{Holder2020,Kaplan2020}, setting it apart in this respect from the length gauge formalism.
	However, there remains the question of how semiclassical results emerge from the perturbative treatment, in other words whether both methods can be brought together as the frequency approaches the limit $\omega \to 0$. This is of course of great theoretical interest; nevertheless, also recent experimental work \cite{Otteneder2020,Burger2019} on materials with resonant frequencies which overlap with the semiclassical regime underscore the need for a better understanding of this crossover regime. 
	
	In this work, we expand on our earlier results, showing that the source of the problems detailed above is the order of limits itself, i.~e. the correct result is obtained by implementing the subsequent limits $\omega \to 0, \tau \to \infty$, which has to date not been rigorously examined. This subtlety probably went unnoticed beforehand because in linear response the shorthand $\omega\rightarrow\omega+i0^+$ is actually sufficient to connect semiclassic and quantum perturbative expressions~\cite{Parker2019}.
	To be concrete, we will explore the \nd~order nonlinear optical response. By using the diagrammatic approach, we reduce all operators to the minimal number of band geometric objects necessary to account for the optical conductivity at this order. We then partition the conductivity into gauge invariant terms with similar geometric origins while keeping all lifetimes finite. 
	By using algebraic identities derived for Bloch states, we reduce the expressions into semiclassical terms, with Fermi surface and Fermi sea contributions which differ in their lifetime dependence. We then demonstrate the connection to the high frequency regime, by showing that the cancellations (and lifetime dependence) seen at low frequencies is due to the interplay of the known shift and injection currents. 
	It turns out that these resonant terms combine with opposite sign as the frequency is reduced. Our formalism not only recovers all previously reported semiclassical contributions, but also allows us to investigate a new term in the conductivity which goes beyond the semiclassical limit. 
	Our derivation introduces several new concepts related to Bloch band geometry which are the result of an induced algebra over the space of single-particle Bloch wavefunctions beyond the traditional Berry curvature.
	
	This paper is organized as follows: In Sec. \ref{sec:prelim} we detail the basic ingredients for nonlinear optical response. Namely, Sec. \ref{sec:fullcond} introduces the nonlinear conductivity at \nd~order in an applied electric field, followed by Sec. \ref{sec:lifetimes} where we describe how to treat finite lifetimes. Sec. \ref{sec:opexp} expands the operators appearing in the conductivity in a suitable form for the subsequent derivation of cancellations. 
	%The effect of the lifetimes prescription on a generic calculation 
	We present the main results of our work \markup{in} Sec. \ref{sec:zerofreq}, where the zero frequency conductivity is derived within the quantum perturbative framework. We also show how cancellations appear at finite frequency in Sec. \ref{sec:finitefreq}. This is followed by a new expression for the conductivity at order $\tau^{-1}$ going beyond the semiclassics in Sec. \ref{sec:beyondsemi}. Finally, consequences of these results are discussed in Sec. \ref{sec:discuss}.
	%%% more references, Cheng, Sipe, Passos, to show the necessity for the amount of discussion we have here
	\section{Preliminaries}
	\label{sec:prelim}
	\subsection{Conductivity at second order}
	\label{sec:fullcond}
	Our starting point is the current produced at \nd~order in perturbation theory, by an applied electric field. This reads: $j^c(\bar{\omega}) = \Re{\sigma^{ab;c}(\bar{\omega}, \omega_1,\omega_2) E_a (\omega_1)E_b (\omega_2)}$, where $\bar{\omega} = \omega_1+\omega_2$, and $\sigma^{ab;c}$ is the \nd order conductivity tensor and $j^c$ is the current density in the $c$ direction. Note that since $\sigma^{ab;c}$ has three spatial components, it is naturally odd under inversion symmetry, and therefore (letting $\mathcal{P}$ be the inversion operator), $\mathcal{P}^{-1} \sigma^{ab;c} \mathcal{P} = -\sigma^{ab;c}$. For there to be a net nonzero \nd order conductivity, inversion symmetry must be absent in the physical system under consideration. However, in what follows, we focus only on $k$-local properties of the response, without applying non-local symmetry restrictions. Our results do not depend on either inversion or time-reversal symmetries. Likewise, the analysis holds for both two- and three dimenionsal systems, unless noted otherwise. In the model system we construct for numerical calculations, we lift both symmetries in order to obtain a finite value for the conductivities when averaging quantities over the Brillouin zone (BZ).
	Focusing on monochromatic perturbations ($|\omega_1|  = |\omega_2|$), two response signals may appear, one for which $\omega_1 = - \omega_2 = \omega$, and the other at $\omega_1 = \omega_2 = \omega$. The first is termed the photovoltaic response while the latter generates a signal at the 2nd harmonic i.e., $\bar{\omega} = 2\omega$. In general, $\sigma^{ab;c}(\bar{\omega})$ can possess both imaginary and real parts. However, starting \markup{from a perturbation of the form} $\vec{E} = \vec{E}_0 \cos(\omega t)$ constrains the tensor to be purely real for both the photogalvanic and 2nd harmonic parts, for linear-polarized light. 
	% The case of circularly polarized light, which antisymmetrizes the response with respect to the spatial indices of the incident perturbation is discussed separately in App. \ref{app:circpol}.
	Importantly, all processes at \nd order contain resonances (absorption or de-excitation) which originate from two distinct processes: $\omega_{1,2}$, which is the result of interaction with a single photon, and $\bar{\omega}$ a process which depends on the sum of frequencies, $\omega_1 + \omega_2$. \markup{Here, $\omega_{1,2}$ refers interchangeably to the two frequencies of the electric field at \nd ~order}. This can be immediately generalized to all other orders, and thus we will have $\omega_1, \omega_1+\omega_2, \omega_1+\omega_2+\omega_3, \ldots, \sum_n \omega_n$ $n$ possible resonances, where $n$ is the order in the expansion. 
	Derived previously from the diagrammatic formalism \cite{Holder2020, Parker2019}, the \nd~order conductivity tensor in a clean system reads,
	\begin{align}
		\sigma^{ab;c}&(\bar\omega,\omega_1,\omega_2) = 
		\notag\\&=
		\frac{-e^3}{\hbar^2}
		\sum_{n}\int_{\bm{k}} \biggl[ \mathcal{U}^{ab;c}_{nn} + \mathcal{W}^{ab;c}_{nn} + \mathcal{V}^{ab;c}_{nn} \biggr]     \label{eq:g2nd} \\ \notag & 
		+ ~~ (a,\omega_1) \lra (b,\omega_2),
	\end{align}
	with the one-loop vertices
	\begin{flalign}
		\mathcal{U}^{ab;c}_{nn} &= \frac{1}{2}f_n \frac{u^{abc}_{nn}}{\left(\omega_1\omega_2\right)}+\frac{1}{2}\sum_{m} \frac{f_{nm}w^{ab}_{nm}v^c_{mn}}{\omega_1\omega_2\left(\bar{\omega} + \vare{n}{m}\right)}&& \label{eq:U}\\ 
		\mathcal{W}^{ab;c}_{nn} &= \sum_m \frac{f_{nm} v^a_{nm} w^{bc}_{mn}}{\omega_1\omega_2(\omega_1 + \vare{n}{m})} &&
		\label{eq:w1} \\ 
		\mathcal{V}^{ab;c}_{nn} &= \sum_{m,l} f_{nm}\biggl(  \frac{v^a_{nm} v^b_{ml} v^c_{ln}}{\omega_1\omega_2\left(\omega_1 + \vare{n}{m}\right)\left(\bar{\omega} + \vare{n}{l}\right)} ~ + ~  &&
		\label{eq:v1}\\ \notag 
		& + \frac{v^c_{nl} v^a_{lm} v^b_{mn}}{\omega_1\omega_2\left(\omega_2 - \vare{n}{m}\right)\left(\bar{\omega} - \vare{n}{l}\right)} \biggr).
		%\label{eq:u}
	\end{flalign}
	Here, $\vare{n}{m} = \varepsilon_n(\bk) - \varepsilon_m(\bk)$, where $\hbar\varepsilon_n(\bk)$ is the energy of Bloch band $n$, at momentum $\bk$, and $\int_\bk = \frac{1}{(2\pi)^d} \int \mathrm{d}^d k$, where $d$ is the number of spatial dimensions in the system. 
	Similarly, with $f_n$ being the Fermi-Dirac  distribution, $f_{nm} = f_n - f_m$. The coupling matrix elements are $v^a_{nm} = \left\langle n (\bk)| \partial_a H | m (\bk) \right\rangle$, $w^{ab}_{nm} = \left\langle n (\bk)| \partial_a \partial_b H | m (\bk) \right\rangle$, $u^{abc}_{nm} = \left\langle n (\bk)| \partial_a \partial_b \partial_c H | m (\bk) \right\rangle$. Throughout, $\partial_a = \frac{\partial}{\partial k_a}$. Here $\left| n(\bk) \right\rangle$ is the cell periodic part of the Bloch wave function for band $n$, i.e. $\left\langle \mathbf{r} | \psi_{n\bk} (\mathbf{r})\right\rangle = e^{i\bk \mathbf{r}}u_{n\bk} (\mathbf{r})$, $u_{n\bk} =\left\langle \mathbf{r} | n(\bk)\right\rangle$.  
	\subsection{Insertion of finite lifetimes}
	\label{sec:lifetimes}
	The expressions derived from perturbation theory necessarily contain resonance factors which are of the form $(\omega_{1,2} \pm \vare{n}{m})^{-l}$, or $(\omega_1 + \omega_2 \pm \vare{n}{m})^{-l}$, where $l >0$, as shown in Eqs.~\eqref{eq:U}-\eqref{eq:v1}. A discussion of the behavior of the conductivity must therefore regulate the divergences introduced by these denominators. As mentioned in the introduction, these lifetimes have a physical origin, such as interactions or disorder effects.
	We show below that the typical prescription, replacing $\omega_{1,2} \to \omega_{1,2} + \frac{i}{\tau}$ and taking immediately the limit $\tau \to \infty$ fails to capture the behavior anywhere outside the immediate vicinity of the resonance; in other words identifying $\frac{i}{\tau} \equiv i0^{+}$ gives erroneous results anywhere outside the resonance. Such a prescription suggests that upon superficial inspection Eqs.~\eqref{eq:U}-\eqref{eq:v1} contain a pole when $\omega_{1,2} \to 0$. We will show this is in fact avoided when the proper replacement is imposed. Ref. \cite{Holder2020} suggested that the diagrammatic formalism allows for extracting the distinction between intra- and inter- band effects, which depend on different lifetimes phenomenologies. Following this reasoning, we implement the following choice which captures all previous suggestions for the insertion of finite lifetimes, 
	\begin{align}
		\omega_{1,2} \to \omega_{1,2} + \frac{i}{\tau}, ~~ \bar{\omega}\to \bar{\omega} + \alpha\frac{i}{\tau}.
		\label{eq:prescrip}
	\end{align}
	Here $\alpha$ is a dimensionless tuning parameter. 
	If $\alpha=1$, one recovers the case where no distinction is made between the higher harmonic ($\bar{\omega}$) and single harmonic ($\omega_{1,2}$) resonances~\cite{Sipe1999,Mikhailov2016}, while $\alpha=2$ corresponds \markup{to the case when the photovoltaic} or \nd~harmonic pole pole at $\omega_1 + \omega_2$ is distinct from the single-photon pole for either $\omega_1, \omega_2$ ~\cite{Passos2018,Parker2019,Holder2020}. We shall detail in the sections below how the value of $\alpha$ determines the semiclassical limit.
	\subsection{Expansion of operators}
	\label{sec:opexp}
	Eqs.~\eqref{eq:U}-\eqref{eq:v1} feature moments of the Hamiltonian which are evaluated by taking their expectation value with respect to the unperturbed eigenstates. To establish the connection with semiclassical theory, we first reformulate these elements using a minimal set of parameters which encode information about the response of the system. This set of these objects is easily inferred by inspection of Eqs.~\eqref{eq:U}-\eqref{eq:v1} and dimensionality arguments. 
	Firstly, in three dimensions, the length dimension at $n$-th order in the electric field of the conductivity $(\omega_1\omega_2\ldots)\hbar^{n-1} e^{-n-1}\sigma^{a_1 a_2 a_3 \ldots; c}$ must be $L^{n+1}$. 
	Energy dimensions enter the expressions through $\varepsilon_n(\bk)$, while length dimensions are obtained from taking momentum derivatives of the cell-periodic part of the wavefunction $\Lambda^{a_1 a_2 ...}_{nm}= \frac{i}{2}\left\langle n(\bk) | \partial_{a_1}\partial_{a_2}\partial_{a_3}\ldots m(\bk)\right\rangle + \textrm{h.c.}$. Since these derivatives are not gauge invariant by themselves, they always appear in connection with lower order objects. This constrains the largest derivative of the wave function to be at most of order $n$, one order less than dictated by power counting. For 2\textsuperscript{nd} order in the applied field we denote the highest order derivative \markup{(\nd~order)} by $\lambda^{ab}_{nm}$~\cite{Kaplan2021a}.
	The objects that will appear in the \nd-order response which encode wave function properties are thus
	\begin{flalign}
		\label{eq:berryconn} 
		r^a_{nm} &= i \left\langle n(\bk) | \partial_a m(\bk) \right\rangle, \\
		\label{eq:lambdaconn}  \lambda^{ab}_{nm} &= \frac{i}{2} \left(\left\langle n(\bk) | \partial_a \partial_b m(\bk) \right\rangle - \left\langle \partial_a \partial_b n(\bk) | m(\bk) \right\rangle\right).
	\end{flalign}
	%and their combinations yield gauge invariant expressions.
	Much of the complexity in the expressions for the conductivity result from the mixtures of energy and length dimensions which are introduced through derivatives the dispersion, i.~e. $\partial_{a_1}\partial_{a_2}\partial_{a_3}\ldots\varepsilon_n(\bk)$, for example the matrix elements of the velocity operator is $v^a_{nn} = \partial_a \varepsilon_n (\bk)$. Since $\varepsilon_n(\bk)$ is a gauge invariant quantity, the largest derivative is of order $n+1$.
	These elements appear in products (specifically, Hadamard products) with the geometrical objects $r^a_{nm}, \lambda^{ab}_{nm}$. For example, $(\varepsilon r^a)_{nm} = \varepsilon_{nm} r^a_{nm}$. Since the conductivity must be gauge invariant, $(\varepsilon r^a)_{nm}$ may not appear by itself, but becomes Hermitian and gauge invariant only in the adjoint form $[A,B]$. It is therefore useful to define a generalized product of the kind $\tlo^{ab,N}_{nm}$~\cite{b_Kaplan2021}, 
	\begin{align}
		\tlo^{ab,N}_{nm} =\left\lbrace \begin{matrix}
			i[\varepsilon^{N} r^a, r^b]_{nm}, 
			& \quad N \text{ even} \\ 
			[\varepsilon^{N} r^a, r^b]_{nm}, 
			& \quad N \text{ odd}
		\end{matrix}\right.
	\end{align}
	Here, the evaluation of commutators is done respecting the matrix structure in the band indices. For example, for $N=0$, $\tlo^{ab,0}_{nm} = i[r^a,r^b]_{nm} = i\sum_{l \neq n,m} (r^a_{nl} r^b_{lm} - r^b_{nl}r^a_{lm}) + i(r^b_{nm}\delta^a_{nm} - r^a_{nm}\delta^b_{nm})$, such that $\delta^a_{nm} = r^a_{nn} - r^a_{mm}$. Note that $\tlo^{ab,0}_{nm}$ is the familiar (non-abelian) Berry curvature. Armed with this intuition, we expand all operators in Eqs.~\eqref{eq:U}-\eqref{eq:v1} such that only $r$, $\lambda$, $\varepsilon$ and $\partial_k \varepsilon$ appear. To this end, recall the definition of the covariant derivative  $\mathbf{D}_\bk = \partial_\bk + i[\cdot,\mathbf{r}]$~\cite{Passos2018,Parker2019} on operators in the Bloch basis:
	%Employing standard notation, all subsequent definitions follow from resolving moments of the Hamiltonian in momentum space. 
	Letting $\mathbf{D}_\bk H_{nm}(\bk) = \partial_\bk H_{nm} + i[H,\mathbf{r}]_{nm}$, the resolution follows by recursively applying this derivative starting with the aforementioned equation. The vertices appearing in the velocity gauge thus become,
	\begin{align}
		u^{ab;c}_{nn} &= [r^a,\Delta^c r^b]_{nn} + [r^a,\Delta^b r^c]_{nn}  + [r^b,\Delta^a r^c]_{nn}  ~ - ~  \notag \\ 
		&\qquad[\varepsilon r^c, \lambda^{ab}]_{nn}  - [\varepsilon r^b,\lambda^{ac}]_{nn} - [\varepsilon r^a, \lambda^{bc}]_{nn} ~ +  ~ 
		\notag \\     &\qquad 
		\frac {i}{2} \left([r^b, \tlo^{ac,1}]_{nn} + [r^a, \tlo^{bc,1}]_{nn}\right)+
		% \notag \\      & \qquad 
		\partial_a \partial_b \partial_c \varepsilon_n,     \label{eq:uu}\\
		w^{ab}_{nm} &= i (\Delta^a_{nm} r^b_{nm} + \Delta^b_{nm} r^a_{nm})
		- \tfrac{1}{2} \left(\tlo^{ab,1}_{nm} + \tlo^{ba,1}_{nm}\right)
		+  \notag\\    &\qquad
		i \varepsilon_{nm} \lambda^{ab}_{nm}+
		%  \notag\\&\quad
		\partial_a\partial_b \varepsilon_n \delta_{nm},     \label{eq:ww}\\
		%\end{align}
		%\begin{align}
		v^a_{nm} &= i\vare{n}{m}r^a_{nm} + \partial_a \varepsilon_n \delta_{nm}.    \label{eq:vv}
	\end{align}
	We defined $\Delta^a_{nm} = v^a_{nn} - v^a_{mm} = \partial_a \vare{n}{m}$.  
	Observe that the objects appearing in the velocity gauge satisfy spatial exchange symmetries which are the result of the smoothness of $H(\bk)$. We therefore pause to examine the symmetry properties of the object $\tlo^{n,ab}$, since it appears, for example, in $w^{ab}_{nm}$, where $a \lra b$ is implied.
	Examining the the behavior of $\tlo^{ab,1}_{nm}$ under the exchange $(a \lra b)$ we find,
	\begin{align}
		\notag \tlo^{ab,1}_{nm} 
		&= \sum_{l} \left(\varepsilon_{nl} r^a_{nl} r^b_{lm}-\varepsilon_{lm} r^b_{nl} r^a_{lm}\right) \\ 
		&=\sum_{l } \left((\vare{n}{m}+\vare{m}{l}) r^a_{nl} r^b_{lm}-(\vare{n}{m}+\vare{l}{n}) r^b_{nl} r^a_{lm}\right) \notag \\
		&=-i \vare{n}{m} \tlo^{ab,0}_{nm} + \tlo^{ba,1}_{nm}.
		\label{eq:permute_tlo1}
	\end{align}
	\allowdisplaybreaks{
		The result is the addition of the object $\tlo^{ab,0}_{nm}$, of an order lower than our starting point, $\tlo^{ab,1}$. In general, a straightforward discussion of the exchange properties of $\tlo^{ab,N}$ is only possible whenever $N \ge 0$, as the case of $N < 0$ involves energy denominators whose exchange is less amenable to simple replacements. Nevertheless, Jacobi identities from the emergent algebra still allow for the derivation of useful identities regarding spatial commutation properties of $\tlo^{ab,N}$.
		Evidently, the structure of $\tlo^{ab,1}$ implies that it only remains symmetric under exchange of spatial components \markup{when considering} its abelian part, i.e., when $n = m$. These objects also bear physical significance. For $a \neq b$, $\tlo^{ab,1}_{nn}$ describes the expression for the orbital magnetization contribution to transport, as it is equal to $\tlo^{ab,1}_{nn} =\sum_{n \in \textrm{occ.}}^{m\in \textrm{unocc.}} \varepsilon_{nm} r^a_{nm}r^b_{mn}$. Generalized curvatures with the energy denominator inverted, such as $\tlo^{ac,-1}_{nn} = \left[\varepsilon^{-1}r^a, r^c\right]_{nn}$ are directly related to corrections to the dipole coupling of the electric field, since $\delta r^c_{nn} = \left[\varepsilon^{-1} r^a, r^c\right]_{nn}$, which are gauge invariant, as shown in the semiclassical formalism \cite{Gao2020}.
		The non-abelian elements which enter whenever interband transitions are considered, are symmetric in spatial indices only up to an addition of the Berry curvature $\Omega^{ab}_{nm}$. Similar identities with the addition of $N-1$-order curvatures $\tilde{\Omega}^{ab,N-1}$ appear at higher orders in the response.}
	\section{Zero frequency expansion}
	\label{sec:zerofreq}
	Our formulation above allows us to expand all the terms appearing for the nonlinear conductivity in a straightforward manner. We begin with the photovoltaic effect, although in the limit $\omega_{1,2} \to 0$, the $\bar{\omega} = 2\omega$ conductivity matches the $\bar{\omega} = 0$ case exactly, by definition. At the limit of $\omega_{1,2} \to 0$, the only remaining parameter in which it is possible to expand is $\tau$. By power counting, we find that the highest power attainable at this order in perturbation theory in the applied electric field is $\tau^2$. We designate every conductivity by $\sigma^{ab;c}_{\tau^n}$, such that $\sigma^{ab;c}_{\tau^n} \propto \tau^{n}$. 
	The full expression for the conductivity $\sigma^{ab;c}(0,0,0)$ is presented in App. \ref{app:D}, after implementing the lifetime insertion prescription of Eq.~\eqref{eq:prescrip}. All lifetimes are kept finite, and a series expansion in $\tau$ is subsequently carried out. The leading order term with $\tau^2$ is, 
	\begin{flalign}
		\notag  \sigma^{ab;c}_{\tau^2} &= -\frac{e^3}{\hbar^2} \tau^2 \int_\bk \sum_n f_n \biggl\{i[r^a, w^{bc}]_{nn} 
		+\frac{i}{2}[r^c, w^{ab}]_{nn}&& \\ \notag
		&\quad  + \frac{1}{\alpha} [r^a \Delta^c, r^b]_{nn} -\frac{1}{2} [r^a,\Delta^b r^c]_{nn}  
		-\frac{1}{2} [r^c,\Delta^a r^b]_{nn}
		&& \\ \notag &\quad 
		+\frac{1}{2} u^{abc}_{nn}
		-\frac{1}{2}\biggl(i[r^b, \varepsilon \tlo^{ca,0}]_{nn} + i [r^c ,\tlo^{ba,1}]_{nn} 
		&& \\ & \quad 
		+ i[r^a, \varepsilon\tlo^{cb,0}]_{nn}
		+ i[r^c, \tlo^{ab,1}]_{nn} \biggr) 
		\biggr\} + (a \lra b). \label{eq:sigma0_res}
	\end{flalign}
	The reduction of this expression relies on the identities established in Eqs. \eqref{eq:uu}-\eqref{eq:vv}. In general, these expressions would contain elements that are either odd or even under the intrinsic permutation symmetry $(a \lra b)$. We can reduce all expressions to objects that are explicitly even/odd under this symmetry. 
	Starting with the first terms in Eq.~\eqref{eq:sigma0_res}, we apply the definition,
	\begin{align} 
		\frac{1}{2} u^{ab;c}_{nn} = \frac{1}{2}  \partial_c w^{ab}_{nn} + \frac{1}{2}  i[w^{ab}, r^c],
	\end{align} 
	thereby removing $u^{abc}$ from Eq.~\eqref{eq:sigma0_res}. The remaining contributions of this term are, 
	\begin{align}
		\notag & \frac{1}{2} \partial_c w^{ab}_{nn} = \frac{1}{2}\partial_a\partial_b\partial_c \varepsilon_n - \frac{1}{2}\partial_c \tlo^{ab}_{nn} = \frac{1}{2} \partial_a\partial_b\partial_c \varepsilon_n + \\ & \notag  \frac{1}{2} [r^b,\Delta^c r^a]_{nn} -  \frac{1}{4} [\varepsilon r^b, \Omega^{ca}]_{nn} \\  -&\frac{1}{4} [\varepsilon r^a, \Omega^{cb}]_{nn} -  \frac{1}{2} [\varepsilon r^a, \lambda^{bc}]_{nn} -\frac{1}{2} [\varepsilon r^b, \lambda^{ac}]_{nn}. 
	\end{align}
	Naturally, this expression retains the intrinsic symmetry $(a\lra b)$ originally possessed by $w^{ab}$. This symmetry is lifted with contribution from the triangle diagram, i.e., those that appear in Eq.~\eqref{eq:vv}. Since $r^a, \varepsilon r^a$ are either Hermitian, or anti-Hermitian matrices, they form a Lie algebra. The triple commutators in Eq.~\eqref{eq:sigma0_res} are resolved using Jacobi identities, \begin{flalign}
		&-[r^c, \tlo^{ba,1}]_{nn}  = - [r^b, \varepsilon\tlo^{ac,0}]_{nn} - [r^a, \tlo^{bc,1}]_{nn}, && \\ 
		&-[r^c, \tlo^{ab,1}]_{nn}  = - [ r^a, \varepsilon \tlo^{bc,0}]_{nn} - [r^b, \tlo^{ac,1}]_{nn}.
	\end{flalign}
	These identities reflect the relationship between the $\tlo^{ab,n}$ and $\tlo^{ab,n-1}$. At this order, their utility is in the fact that while $\tlo^{ab,1}_{nm}$ is by itself neither even nor odd under $a \lra b$, $\tlo^{ab,0}_{nm}$ is always odd under this symmetry. The Jacobi identities furnish a separation of the pieces in $\sigma_{\tau^2}^{ab;c}$ into terms which are strictly even or odd under the intrinsic permutation symmetry. Combining these identities with the remaining terms yields,
	\begin{flalign}
		\sigma_{\tau^2}^{ab;c} &= - \frac{e^3}{\hbar^2} \tau^2 \int_\bk \sum_n  f_n\biggl\{
		\frac{1}{2}\partial_a\partial_b\partial_c \varepsilon_n 
		+ \frac{1 - \tfrac{2}{\alpha}}{2} [r^a,\Delta^c r^b]_{nn} 
		&& \notag \\ \notag
		&\quad 
		- \frac{1}{4} [ r^a, \varepsilon \Omega^{cb}]_{nn} 
		+  \frac{1}{4} [ r^b, \varepsilon \Omega^{ca}]_{nn}
		&&  \\ \notag 
		&\quad
		-\frac{1}{2} [\varepsilon r^b, \lambda^{ac}]_{nn} + \frac{1}{2} [\varepsilon r^a, \lambda^{bc}]_{nn}
		\biggr\} + (a\lra b) 
		\\ 
		&  = - \frac{e^3}{\hbar^2} \tau^2 \int_\bk \sum_n  f_n
		\Bigl(
		\partial_a\partial_b\partial_c \varepsilon_n + [r^a,\Delta^c r^b]_{nn}\left(1 - \tfrac{2}{\alpha}\right)
		\Bigr).
		\label{eq:tau2}
	\end{flalign}
	Continuing with the expansion, the terms at order $\tau^1$ are
	\begin{widetext}
		\begin{flalign}
			\sigma_{\tau}^{ab;c} &= -\frac{e^3}{\hbar^2}\tau \sum_{n}
			f_n\biggl\{\int_\bk \frac{i \alpha}{2} \left[\frac{r^c}{\varepsilon}, w^{ac} \right]_{nn}-i  \left[\frac{r^a}{\varepsilon}, w^{bc}\right]_{nn}  ~ + ~  \frac{(1+\alpha)}{2i} \left[\frac{r^a}{\varepsilon}, \Delta^b r^c\right]_{nn} - \frac{(1+\alpha)}{2i}  \left[\frac{r^c}{\varepsilon}, \Delta^a r^b\right]_{nn}  + \notag && \\ 
			&\quad -\frac{i}{2}\left[\frac{r^a}{\varepsilon},\tlo^{bc,1}\right]_{nn}  -\frac{i\alpha}{2}\left[r^a,\varepsilon\tlo^{cb,-1}\right]_{nn} \notag + \frac{(1+\alpha)}{2i}\left[r^a,\tlo^{bc,0}\right]_{nn} \biggr\}+ (a \lra b) 
			\notag  \\ 
			& = -\frac{e^3}{\hbar^2} \tau \sum_n \int_\bk f_n
			\biggl\{
			\partial_b \Omega^{ac,0}_{nn} +
			\left(\tfrac{\alpha}{2}-1\right)\left(\left[\varepsilon r^a, \tlo^{bc,-1}\right]_{nn} + [r^b, \lambda^{ac}]_{nn} \right)\biggr\} ~ + ~ (a\lra b)
			.
			\label{eq:tau1}
		\end{flalign}
	\end{widetext}
	{\allowdisplaybreaks
		%We introduced $\sum_{n \in \textrm{occ.}} = \sum_n f_n $ for brevity. 
		In the last line, for the benefit of familiarity we reintroduced the derivative of the abelian Berry curvature $\Omega^{ac,0}_{nn}$ using the geometric connections Eqs.~\eqref{eq:berryconn}-\eqref{eq:lambdaconn}~\cite{Holder2020},
		\begin{align}
			\partial_b \Omega^{ac,0}_{nn} = -\frac{1}{2} \left[r^b, \Omega^{ac,0}\right]_{nn} + i\left[r^a, \lambda^{bc}\right]_{nn} - i\left[r^c, \lambda^{ab}\right]_{nn}.
			\label{eq:berrydipole}
		\end{align}
		The reduction of Eq.~\eqref{eq:tau1} employed the resolution of $w^{ab}$. $\left [\frac{r^c}{\varepsilon}, w^{ab}\right ]_{nn} = i\left[\frac{r^c}{\varepsilon}, \Delta^a r^b \right] - i[r^c, \lambda^{ab}]_{nn} -\frac{1}{2} \left[\frac{r^c}{\varepsilon}, \tlo^{ab,1}\right]_{nn} + (a \lra b)$. The term $[r^c, \lambda^{ab}]_{nn}$ appears without any band energy factors, as these naturally cancel once the commutator is resolved. We further used the identity $\left[\frac{A}{\varepsilon}, \varepsilon B\right]_{nm} = \varepsilon_{nm} \left[\frac{A}{\varepsilon}, B\right]_{nm} - \left[A,  B\right]_{nm}$. 
		The imposition of the $(a \lra b)$ condition enforces most cancellations for this term. In particular, $f_n \left[\varepsilon^{-1} r^a, w^{bc}\right]_{nn}$ contributes $ f_n [\varepsilon^{-1} r^a, \Delta^b r^c]$ which is also contributed by $\mathcal{V}^{ab;c}_{nn}$. In the case of Hall responses, we shall show that terms of this form cancel identically. We have, $f_n [\varepsilon^{-1} r^a, \Delta^b r^c] = \sum_{n,m} \vare{n}{m}^{-1} r^a_{nm}\Delta^b_{mn} r^c_{mn} = \frac{1}{2} \sum_{nm} f_{nm}\left(\vare{n}{m}^{-1} \Delta^b_{mn} \left(r^a_{nm} r^c_{mn} - r^a_{mn}r^c_{nm}\right)\right)$. the term within the brackets is clearly purely imaginary, which coupled with the requirement that overall zero frequency response be real, vanishes. This extends to all constituents of $\mathcal{W}^{ab;c}_{nn}$ and $\mathcal{V}^{ab;c}_{nn}$.}
	Additionally, in the expansion of $\mathcal{V}^{ab;c}_{nn}$ one encounters terms of the following composition, $2 f_{nm} r^a_{nm} \frac{\varepsilon_{ml}}{\varepsilon_{nm}}  r^b_{ml} r^c_{ln} + \textrm{c.c.}$. Here, the factor of two is broken apart to yield two terms: $\left[\varepsilon^{-1} r^a,\left[\varepsilon r^b, r^c\right]\right]$ and $-\left[r^a,\left[r^b, r^c\right]\right] + \left[\varepsilon^{-1} r^a,\left[r^b, \varepsilon r^c\right]\right]$. \markup{The latter here can be immediately summed} with the first, resulting in $\left[\varepsilon^{-1} r^a, \tlo^{bc,1}-\tlo^{cb,1}\right] = -[r^a, [r^b, r^c]] = -\left[r^a, \tlo^{bc,0}\right]$, using Eq.~\eqref{eq:permute_tlo1}. In this manner, identities related to the fundamental permutation symmetries of the generalized Berry curvatures enter ubiquitously at every order. The emergence of $\left[r^a, \tlo^{bc,0}\right]$ allows for the interpretation of the resulting terms, which when summed together give Eq.~\eqref{eq:berrydipole}. 
	The third term is $\tau^0$. Substituting the resolution of $w^{ab}$ explicitly, we find:
	\begin{widetext}
		\begin{align}
			\notag \sigma_{\tau^0}^{ab;c} &= -\frac{e^3}{\hbar^2} \sum_{n} \int_\bk f_n\biggl\{
			\frac{1}{\alpha}\left[\varepsilon^{-2} r^a, \Delta^c r^b \right]_{nn} +  
			(1-\alpha)^2\left[\varepsilon^{-2} r^c, \Delta^a r^b \right]_{nn} -  \alpha^2 \left[r^a, \varepsilon \tlo^{cb,-2}\right] \frac{\alpha^2}{2} \left[r^a,\tlo^{cb,-1}\right]_{nn} \\ 
			&\quad -\notag  \alpha  \left[ r^a,  \tlo^{cb,-1}\right]_{nn} + \alpha  \left[ r^a,  \varepsilon^{-1}\tlo^{cb,0}\right]_{nn} + i \left[ r^a,  \varepsilon^{-1}\tlo^{cb,0}\right]_{nn} ~+~  i \left[ r^a,  \varepsilon^{-1}\tlo^{cb,0}\right]_{nn} + \left[ r^a,  \varepsilon^{-2}\tlo^{cb,1}\right]_{nn}\\ 
			&\quad+ \notag  \alpha^2 \left[r^c, \varepsilon \tlo^{ab,-2}\right]_{nn} + \frac{\alpha^2}{2}  \left[r^c,\tlo^{ba,-1}\right]_{nn}  - \alpha  \left[ r^c,  \tlo^{ba,-1}\right]_{nn} -   \left[ r^c,  \varepsilon^{-2}\tlo^{ab,1}\right]_{nn} +  \frac{\alpha^2}{2}  i\left[\varepsilon^{-2} r^c, w^{ab}\right]_{nn}  \\  
			&\quad + ~ i\left[\varepsilon^{-2} r^a, w^{bc}\right]_{nn} \biggr\}+ (a \lra b) 
			\notag\\
			&= -\frac{e^3}{\hbar^2} \sum_n \int_\bk f_n \biggl\{\partial_c \tlo^{ab,-1}_{nn} - \frac{1}{2}\partial_a \tlo^{bc,-1}_{nn}  + \eta^{ab;c}_{nn} + H^{ab;c}_{nn} (\alpha,k)\biggr\} + (a \lra b),
			\label{eq:tau0}
		\end{align}
	\end{widetext}
	where we defined,
	\begin{align}
		\notag &\eta^{ab;c}_{nn} = \frac{i}{2} \biggl(\left[r^a, \tlo^{bc, -1}\right]_{nn} + \left[\varepsilon^{-1} r^a, S^{bc}\right]_{nn}  - \left[\varepsilon^{-2} r^a, \Delta^c r^b \right]_{nn} \\ & \qquad - (a \leftrightarrow c) \biggr) \\ 
		&H^{ab;c}_{nn} = \sum_n f_n \frac{\alpha-2}{2} \biggl(i\left[r^c,\tlo^{ba,-1}\right]_{nn} + i\left[ r^a,  \tlo^{cb,-1}\right]_{nn} \notag \\
		& \quad - \left[\varepsilon^{-2} r^a, \Delta^c r^b \right]_{nn} -  (a \lra c)\biggr).
	\end{align}
	and introduced the spatially symmetric, gauge covariant interband tensor $S^{ab}_{nm} = -\frac{i}{2} r^a_{nm} \delta^b_{nm} +\frac{1}{2} \lambda^{ab}_{nm} + (a \leftrightarrow b)$ which in generalization of the (positional) shift vector~\cite{Sipe2000} can be viewed as a shift in area.
	While the intermediate expressions \markup{grow longer with each descending order} in $\tau$, the principle guiding the reduction process, and the ubiquity of Jacobi identities remains unchanged. At this order, we used the identity
	\begin{align}
		[r^a, \lambda^{bc}] + (a \lra b) + (a \lra c) = 0,
	\end{align}
	which leads to the expansion
	\begin{align}
		& \notag \partial_c [\varepsilon^{-1} r^a, r^b ]_{nn} = \left[\varepsilon^{-1} \lambda^{ac}, r^b\right]_{nn} +\frac{1}{2}\left[ \varepsilon^{-1} \tlo^{ca,0}, r^b\right]_{nn} + \\ &  \left[\varepsilon^{-1} r^a, \lambda^{bc}\right]_{nn} +\frac{1}{2} \left[\varepsilon^{-1} r^a, \tlo^{cb,0}\right]_{nn} +  \left[\varepsilon^{-2} r^a, \Delta^c r^b \right]_{nn}.
	\end{align}
	Eqs.~(\ref{eq:tau2}, \ref{eq:tau1},\ref{eq:tau0}) for respectively $\sigma_{\tau^2}$, $\sigma_{\tau^1}$ and $\sigma_{\tau^0}$ constitute the semiclassical dc-conductivity in the bulk, and is consistent with all previously reported expressions for terms which have been known before~\cite{Sodemann2015,Parker2019,Gao2019,Holder2020}. At the same time, all explicitly $\alpha$ dependent terms have not been discussed so far in the literature and are novel.
	Some general observations may be drawn from these results: The form of all conductivities contains two elements: the first is a total derivative term, directly related to the Fermi surface, which is independent of the lifetime ratio $\alpha$. The other is a combination of geometric terms, which cannot be reduced to a total derivative, and which originate from the Fermi sea. These latter terms are strongly dependent on the ratio $\alpha$, but a value of $\alpha$ exists for which they can be eliminated. The appearance of the $\alpha$ dependent contributions is a signature of processes beyond semiclassics which are accessible only by the diagrammatic method. 
	
\begin{figure*}[t]
		\centering
		\includegraphics[width=0.40\textwidth]{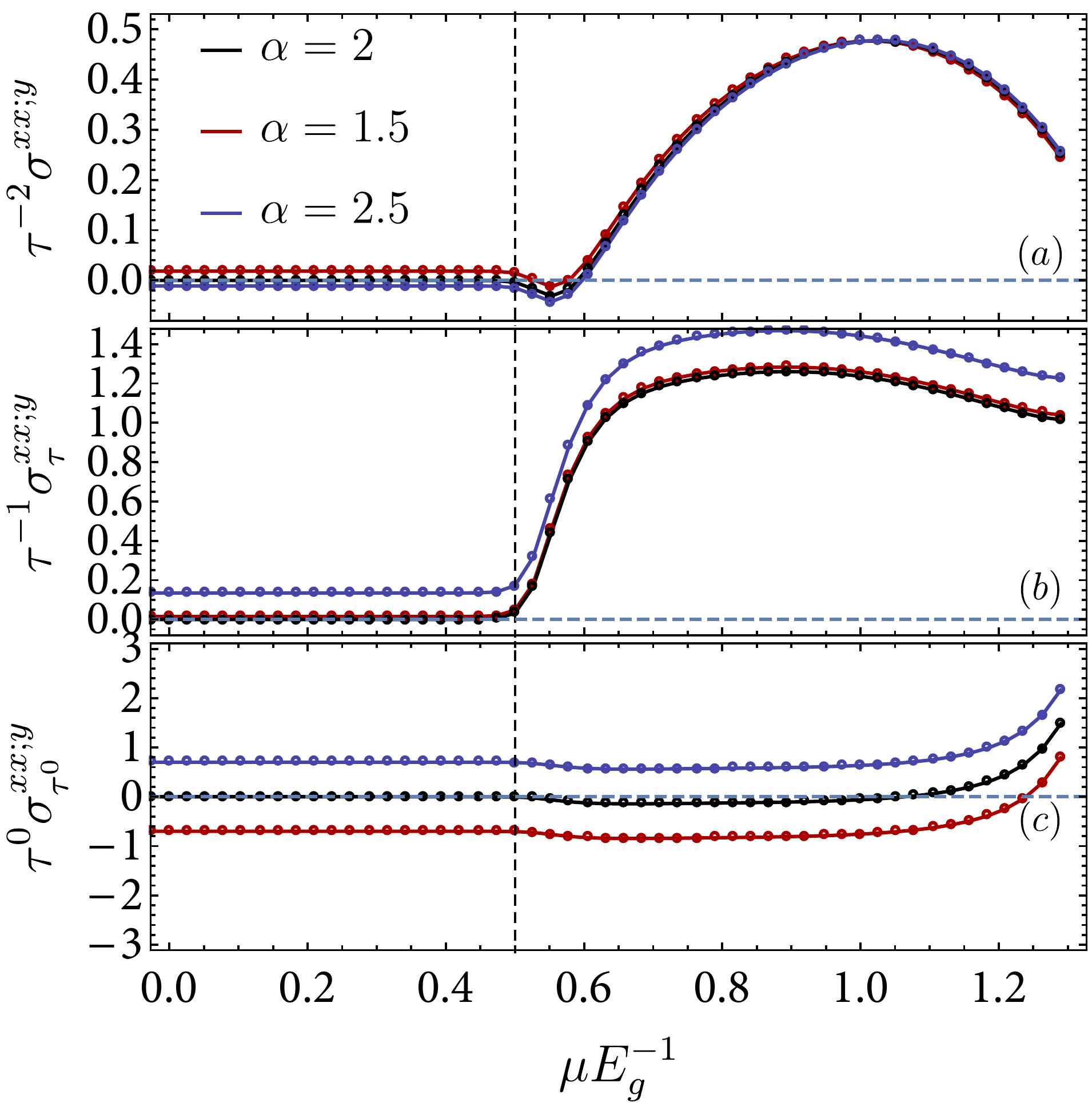}
		\includegraphics[width=0.40\textwidth]{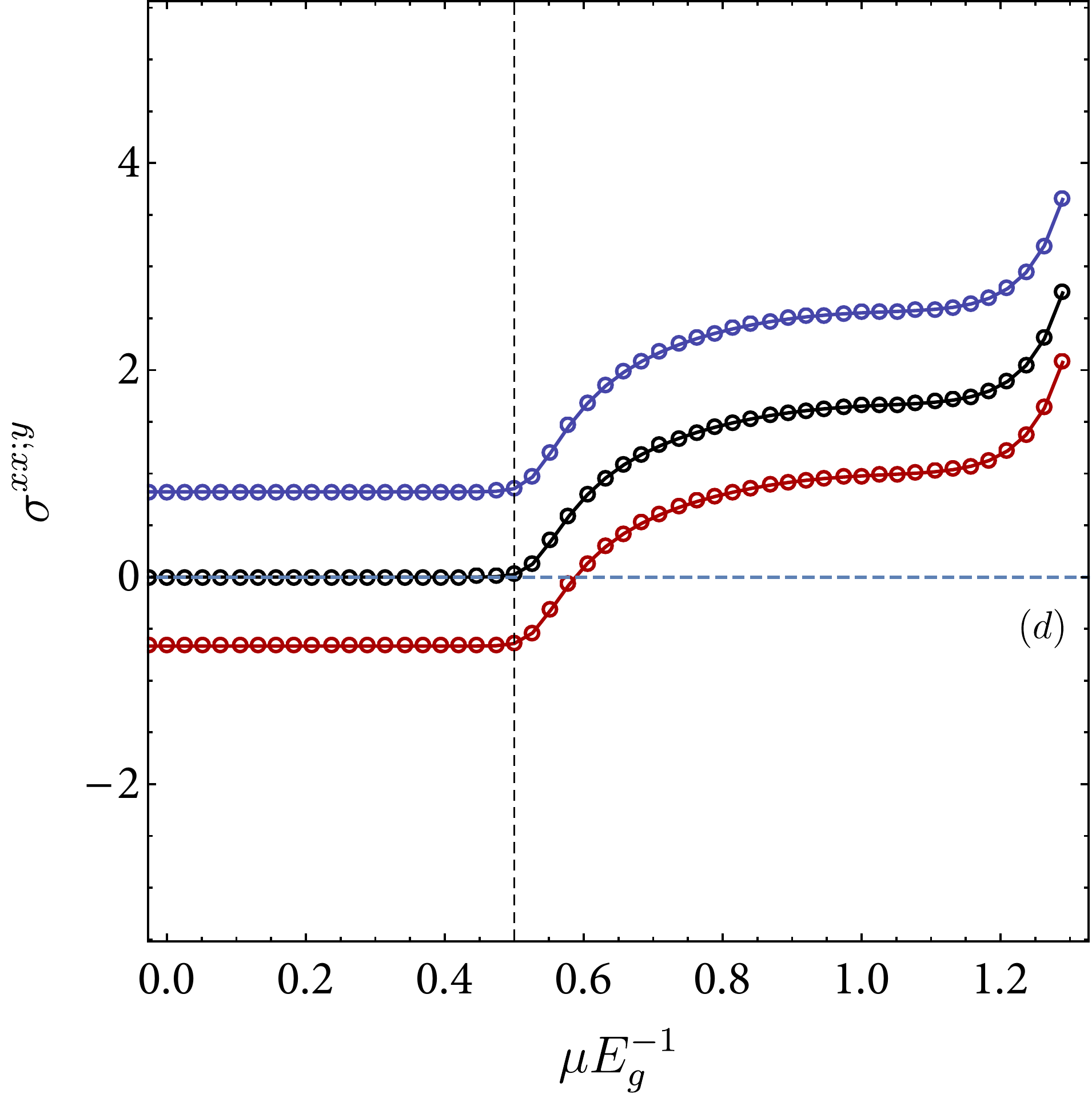}
		\caption{$\sigma_{\tau^2}^{xx;y},\sigma_{\tau^1}^{xx;y},\sigma_{2}^{xx;y}$, shown in (a), (b), (c), respectively as a function of chemical potential $\mu$, for several values of $\alpha = 1.5,2.0,2.5$. In (d), we plot the total conductivity, with $\tau = 1$. The interval in $\mu$ to the left of the vertical dashed lines denotes the gapped region of the model in App. \ref{app:model}. Note that for $\alpha = 2$, the in-gap signal vanishes identically, as all the remaining terms are now require a Fermi surface. \markup{The proper reflection of this Figure} for $\mu < 0$, due to particle-hole symmetry, is explained in App.~\ref{sec:particle_hole}. The lines are guides to the eye, and the azure horizontal dashed lines denotes the zero ordinate for each conductivity $\sigma_{\tau^n}^{xx;y}$. For ease of presentation we set $e = \hbar = 1$.}
		\label{fig:fig4}
	\end{figure*}

	In order to illustrate the effect of all terms on the zero frequency conductivity, we construct a model of a square lattice, with next-nearest neighbor hoppings, which generically breaks time-reversal and inversion symmetries. The model is gapped, and the conductivity for every $\tau^n$ is presented against the value of the chemical potential $\mu$. The system becomes metallic whenever $|\mu| > \frac{E_g}{2}$, where $E_g$ is related to the size of the time-reversal breaking gap term. The full model is presented in App. \ref{app:model}.
	Since all conductivities contain an $\alpha$-dependent part, we additional plot them for several values of $\alpha$. Figs.~\ref{fig:fig4}(a,b,c) present the Hall conductivity $\sigma^{xx;y}$ for $\tau^2, \tau^{1}, \tau^{0}$, respectively (without $\eta^{ab;c}_{nn}$) for the $\alpha$ values $\alpha = 1.5, 2, 2.5$. This conductivity relates the current in the $y$-direction to applied electric field in the $x$-direction. 
	Overall, the effect of the $\alpha$-dependent terms is most significant only when the leading order Fermi-surface term is small. In Fig. \ref{fig:fig4}(a), for example, the additional $\alpha$-dependent contribution is in fact negligible near $\mu \sim 0.8 E_g$, where the leading order nonlinear Drude weight is in fact maximal. Since $\alpha$-dependent contributions are derived from the Fermi sea, they may produce an anomalous signal within the gap of an insulator, which occurs in the model for $|\mu| < \frac{E_g}{2}$. This is seen in all curves (a,b,c), $\alpha \neq 2$ in Fig.~\ref{fig:fig4}. For $\alpha=2$, the in-gap conductivity vanishes all together, leaving only Fermi surface contributions. 
	\section{Finite frequency expressions}
	\label{sec:finitefreq}
	In this section we present a decomposition of the total nonlinear conductivity at \nd~order, based on a resonance analysis. 
	We note that analyses of nonlinear resonant responses is well documented in the literature~\cite{Sipe2000,Young2012,Ventura2017,Passos2018,Parker2019,Holder2020}. Nevertheless, previous works have been restricted to near-resonant behavior and did not or even out of principle cannot characterize the behavior far from resonance.
	
	The motivation for the partitioning of the conductivity which we demonstrate below stems from the understanding that for any finite $\tau<\infty$, resonances cannot vanish identically even in the gap~\cite{Kaplan2020}. Therefore, if some part of the conductivity is non-zero just above gap, it should persist smoothly as the frequency is decreased. 
	On the other hand, in clean systems the dissipative components of the in-gap conductivity are expected to vanish.
	If a resonant term does not vanish by itself, it must therefore be canceled by another piece of the conductivity.
	For such a cancellation to occur, the conductivity must contain elements which have similar magnitudes but enter with opposite signs. 
	%This follows the structure of the cancellations and identities of geometric objects derived in Sec. \ref{sec:zerofreq}. 
	A finite $\tau$ also implies that the eigenenergy surface of the resonance is naturally broadened by $\sim \tau$ which suggests that Eqs.~\eqref{eq:U}-\eqref{eq:v1} do not necessarily vanish separately in the gap of an insulator, but only through some combination of their constituents. 
	Indeed, it is known that a non-vanishing signal persists even if TRS imposed on the system, and cancellations are enforced by sum rules~\cite{Sipe1999}. 
	The starting point of our analysis is the different physical origin of Eqs.~\eqref{eq:U}-\eqref{eq:v1}, as seen in the diagrammatic approach. For example, the three legged diagrams $\mathcal{V}^{ab;c}_{nn}$ contain objects which are also represented in $\mathcal{W}^{ab;c}_{nn}$, but since $\mathcal{W}^{ab;c}_{nn}$ is produced from higher-order derivatives in $H(\bk)$, it should be viewed as a counter-term to $\mathcal{V}^{ab;c}_{nn}$~\cite{Mahan1990}. 
	Since all relations will be local in $k$, we impose no symmetry constraints (neither TRS nor inversion symmetry are implied). The only symmetrization carried out explicitly is the imposition of the intrinsic permutation symmetry of the conductivity tensor, i.e., $(a, \omega) \lra (b, -\omega)$, for linear polarized light.
	\subsection{Bulk photovoltaic effect}
	\label{sec:bulkphotovoltaic}
	The bulk photovoltaic effect is obtained when the frequencies of the incoming field are chosen such that $\bar{\omega} = 0, \omega_1 = -\omega_2 = \omega$. In the following we focus on the two regimes relevant at finite frequency. For finite $\tau$, resonances are controlled by the denominator $N_{nm}(\omega) = (\omega - \vare{n}{m} + i \tau^{-1})^{-1}$\markup{,} which emerges from the diagrammatic analysis. At the resonance, i.e., when $|\omega - \vare{n}{m}| \ll \tau^{-1}$, $\Re{N_{nm}} \to 0, \Im{N_{nm}} \sim i\tau$, while off-resonantly $\dO = |\omega - \varepsilon_{nm}| \gtrsim \tau^{-1}$, $\Re{N_{nm}} \sim \dO^{-1}, \Im{N_{nm}} \sim i\dO^{-2} \tau^{-1}$.
	Writing for general $N_{nm}(\omega)$, we have,
	{\allowdisplaybreaks
		\begin{flalign}
			\sigma^{ab;c}_{\textrm{inj}} &= \frac{e^3}{\hbar^2}\frac{\tau}{\alpha} \sum_{nm} \int_\bk f_{nm} N_{nm}(\omega) r^a_{nm} r^b_{mn} \Delta^c_{mn} + (a \lra b) \label{eq:res1} && \\ 
			\notag \sigma^{ab;c}_{\textrm{c-inj}} &= \frac{e^3}{\hbar^2 } \sum_{nm} \int_\bk  f_{nm} N_{nm}(\omega) \biggl\{ -\frac{ r^a_{nm} r^b_{mn} \Delta^c_{mn}}{\vare{m}{n}}  ~ + ~ \notag \\ & r^a_{nm}\lambda^{bc}_{mn} - \frac{r^a_{nm} \tlo^{bc,1}_{mn}}{2\vare{m}{n}}- r^a_{nm} r^c_{mn} \delta^b_{mn}\biggr\} ~+~ (a \lra b) \label{eq:res4}.
	\end{flalign}}
	In our definitions, the prefix $c-$ denotes a counter term to the expression without it. Through this decomposition, it is evident that $c-$ terms reverse in sign relative to their main counterparts. Note that the injection current counter term (Eq.~\eqref{eq:res4}) contains additional geometrical terms. While this may in principle affect the sign of this term, in systems where the injection current is non-zero (for linear polarized light) the contributions of these terms are always subleading compared to the pole contribution. 
	Importantly, all previously known resonant terms are recovered in Eqs. \eqref{eq:res1}-\eqref{eq:res4}. Eq.~\eqref{eq:res1} is the familiar injection current \cite{Holder2020}, while the shift current is recovered exactly in  Eq.~\eqref{eq:res4}. 
	We note that due to the common assumption of TRS, the first term in Eq.~\eqref{eq:res4} has been previously disregarded in the consideration of the shift current~\cite{Sipe2000}, even though this term has the same lifetime phenomenology ($\tau^0$) as the remaining terms. Eqs.~(\ref{eq:res1}-\ref{eq:res4}) are gauge invariant by themselves, and therefore lead to cancellations at every $k$ point, making the current taper off below the gap.
	Note that even below the gap, the various contributions do not necessarily vanish. This is because the off-resonant $\Re{N_{nm}}$ of the denominators survives for finite $\tau$. The prescription whereby $\tau \to \infty$ is taken at this stage \cite{Nastos2006} would lead to \markup{incorrect results} for TRS-broken systems. The resonant part is described, as usual by taking the imaginary of $N_{nm}$, which in the limit of $\tau \to \infty$ reduces to the familiar $\delta(\omega \pm \vare{n}{m})$.  After fixing $N_{nm}$, one may start from the regime $\omega \ll \vare{n}{m}$, with all other terms simply diverging slower than $\omega^{-2}$. Additionally, the Fermi surface (FS), which is distinctly off-resonant, may contribute. Retaining the leading order contribution when $\vare{n}{m} \gg \omega$, we write,
	\begin{align}
		\sigma_{\textrm{FS}}^{ab;c} &= -\frac{e^3}{2 \hbar^2 \omega_1\omega_2} \int_\bk \sum_{nm}  \left(f_n u^{ab;c}_{nn}\delta_{nm} + i f_{nm} w^{ab}_{nm}r^c_{nm}\right) =    \notag  \\
		&  = -\frac{e^3}{2 \hbar^2 \omega_1\omega_2} \int_\bk f_n \left(\partial_a \partial_b \partial_c \varepsilon_n - \partial_c \tlo^{ab,1}_{nn}\right).
		\label{eq:fs}
	\end{align}
	Reading off the prescription in Sec.~\ref{sec:lifetimes}, one makes the substitution $\omega_1 \to -\omega + \frac{i}{\tau}$, and $\omega_1 \to \omega + \frac{i}{\tau}$. In the limit \markup{of $\omega \to 0$}, the FS contribution leads with $ \sim \tau^2$, and is purely real. To cancel the term involving the higher-order curvature $\partial_c \tilde{\Omega}^{ab,1}$, the off-resonant parts are also to be expanded in the limit of finite $\tau$ and fixed $\alpha$. Then, the identities presented in Sec.~\ref{sec:zerofreq} can be applied. A true cancellation of this term and the recovery of the Drude weight as the leading order $\tau^2$ FS contribution \cite{Watanbe2020, Watanbe2021} is only possible if $\alpha = 2$, and the proper lifetime prescription is implemented.
	In order to explain the cancellation between injection and shift current below the gap, we apply the following simple argument. By expanding in powers of $\omega \ll E_g$ the leading order injection contribution scales as $\sigma_{\textrm{inj}} \sim \tau \Im{N_{nm}(\omega)} \sim E_g^{-2} \tau^{0}$. Naturally, for Eq.~\eqref{eq:res4}, the scaling is of the form $\sigma_{\textrm{shift}} \sim \tau^{0} \Re{N_{nm}(\omega)} \sim E_g^{-2} \tau^{0}$. We thus conclude that below the gap, the injection and shift terms acquire similar relative size. But crucially, the sign of the shift current in this scenario is inverted compared to the injection current, as dictated by the cancellations between the $\mathcal{V}^{ab;c}_{nn}$ and $\mathcal{W}^{ab;c}_{nn}$ vertices, derived in Sec.~\ref{sec:zerofreq}. Moreover, in contrast to Ref.~\cite{Kaplan2020}, this argument holds for arbitrary local symmetries of the underlying (e.g., without assuming $\mathcal{P}\mathcal{T}$ symmetry), and is independent of spatial direction in which the current is probed, relative to the applied field (Ref.~\cite{Kaplan2020} considered the longitudinal contribution $\sigma^{xx;x}$). This can also be immediately read off from Eqs.~\eqref{eq:res1}, ~\eqref{eq:res4}.
	\begin{figure}[h]
		\centering
		\includegraphics[width=1.0\columnwidth]{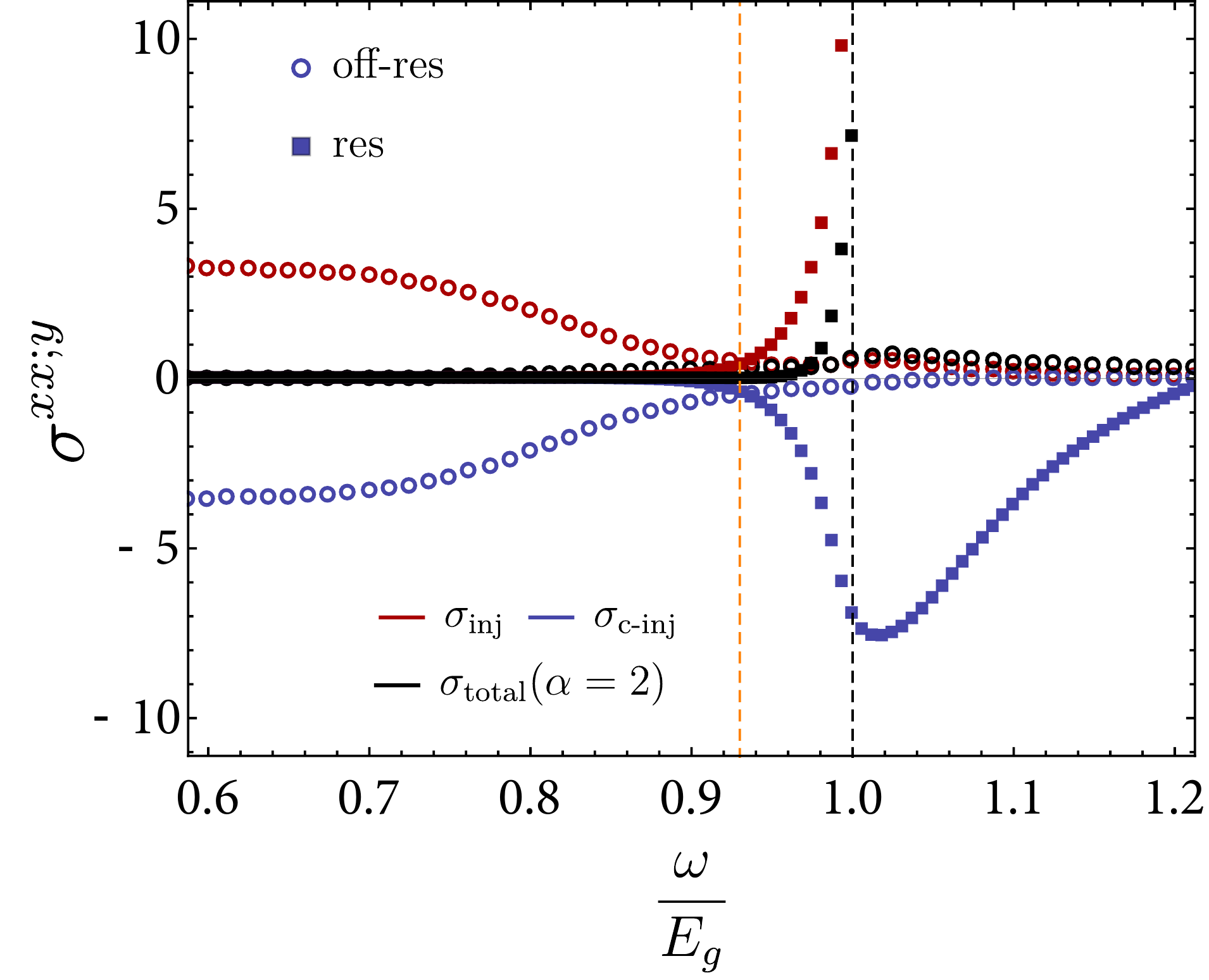}
		\caption{Decomposition of the total conductivity, as a function of $\omega E_g^{-1}$, for the model in App.~\ref{app:model}. The dashed gray line indicates the position of the gap. Filled squares (of any color) denote resonant contributions, empty circles (of any color) denote off-resonant contributions. In red, we plot the injection current contribution, Eq.~\eqref{eq:res1}. In blue, we plot the counter-terms, which are the Eq.~\eqref{eq:res4}. Black denotes the total optical conductivity. Throughout, the chemical potential is placed within the gap, i.e., $\mu = 0$, and Eq.~\eqref{eq:fs} vanishes identically.}
		\label{fig:fig1}
	\end{figure}
	In Fig.~\ref{fig:fig1} we plot a numerical integration of $\sigma_{\textrm{inj}}^{ab;c}$,  $\sigma_{\textrm{c-inj}}^{ab;c}$, based on a lattice model defined in App. \ref{app:model}. Clearly, the injection contribution (red) and counter-term (blue) are opposite in sign. We further separate the two contributions to the total conductivity which are either resonant and off resonant. One can observe that as expected $\sigma_{inj}$ enters the conductivity with a sign opposite to that of the counter term, $\sigma_{\textrm{c-inj}}$. We define all resonant terms as \markup{$\sigma_{\textrm{res}} \sim \Im{N_{nm}(\omega)}$, while the off-resonant contributions are obtained when $\sigma_{\textrm{off-res}} \sim \Re{N_{nm}(\omega)}$}.
	In Fig. \ref{fig:fig1} we denote the onset of the semiclassical regime by an orange vertical line. This point is defined by the intersection of the resonant terms, $\sigma_{\textrm{res}}$, with that of the off-resonant terms $\sigma_{\textrm{off-res}}$. The position of this point can be estimated by expansion in the vicinity of the resonance and equating both terms. To leading order in $\dO = \omega - \vare{n}{m} \approx \omega - E_g$, we find
	\begin{align}
		\dO = \frac{2 E_g}{\alpha \tau}\frac{\int_\bk \sum_{nm} f_{nm} \vare{m}{n}^{-1} r^a_{nm} r^b_{mn} \Delta^c_{mn}}{\int_\bk \sum_{nm} f_{nm}r^a_{nm} r^b_{mn} \Delta^c_{mn}} + \mathcal{O}(\tau^{-2}).
		\label{eq:semiclass}
	\end{align}
	This calculation reinforces the idea that the semiclassical regime and its onset are affected by the lifetime ratio $\alpha$. In Fig.~\ref{fig:fig2}, we show the behavior of the total conductivity (resonant and off-resonant) for other values of $\alpha$. One can observe the emergence of a dip (minimum) as a function of $\alpha$, once $\omega < E_g$. However, only for the value $\alpha = 2$ does the conductivity tend as expect to zero, below the gap. Instead, depending on whether $\alpha$ deviates below are above $2$, one obtains a finite conductivity. The onset of the semiclassical regime, as defined by Eq.~\eqref{eq:semiclass} can also be seen in Fig.~\eqref{fig:fig2}, \markup{as corresponding roughly to the region where} the conductivity reaches an inflection point. However, the behavior, as suggested by Eq.~\eqref{eq:semiclass} is somewhat more intricate. To elucidate the dependence of $\dO$ we evaluate both the exact position (circles) vs the approximated relation (solid line) exhibited by Eq.~\eqref{eq:semiclass}. This is presented in Fig.~\ref{fig:fig2} (inset), which shows the dependence $\dO(\alpha)$.
	\begin{figure}[h]
		\centering
		\includegraphics[width=1.0\columnwidth]{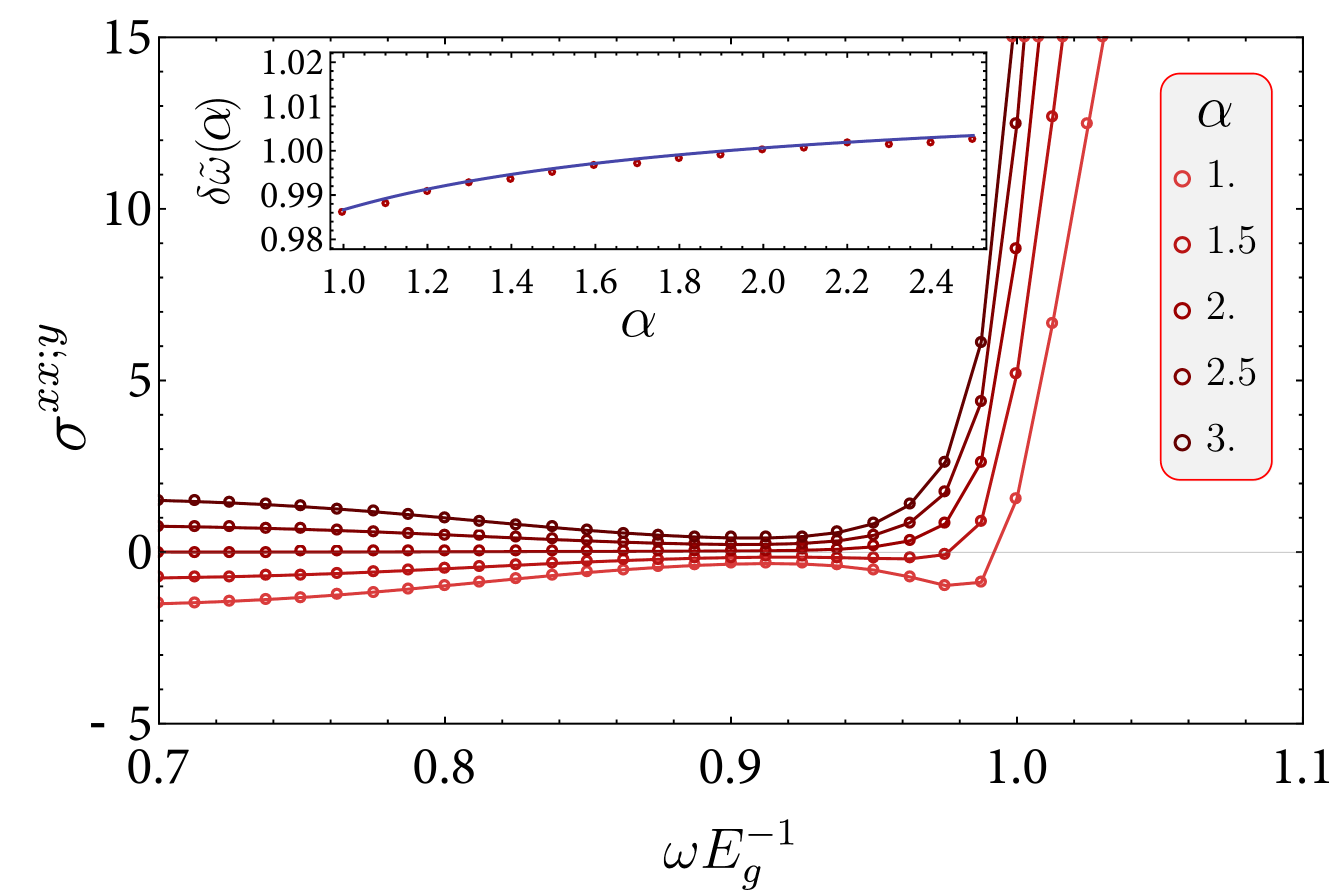}
		\caption{Dependence of the total conductivity $\sigma^{ab;c}$ and the position of the onset of semiclassics as a function of the lifetime ratio $\alpha$. The chemical potential is in the gap, throughout. Total conductivity for different values of the ratio $\alpha$. For $\alpha < 2$, the conductivity flips when crossing below the gap. Inset: $\dO$ normalized to the value at $\alpha=2$. The empty circles denote the numerically precise value showing the onset of semiclassics. The blue line is Eq.~\eqref{eq:semiclass}. The typically adopted $\alpha = 1$ results in an onset of semiclassics earlier, that is closer to the gap, than at $\alpha=2$.}
		\label{fig:fig2}
	\end{figure}
	{\allowdisplaybreaks
		\subsection{2\textsuperscript{nd} harmonic response} 
		A \markup{monochromatic} perturbation also generates a response at $\bar{\omega} = 2\omega$. Compared to the photovoltaic effect, however, there is no distinct scale separation in the lifetime $\tau$ in the terms constituting the conductivity. In the photovolatic limit, the injection current (Eq.~\eqref{eq:res1}), for example, is distinguishable from all other processes since it scales like $\sim \tau$, whereas all other terms (Eq.~\eqref{eq:res4}) are of order $\tau^{0}$. However, from the analysis in Sec.~\ref{sec:zerofreq}, it is also apparent that when $\omega \to 0$, $\sigma^{ab;c}(2\omega, \omega,\omega)$ must tend to $\sigma(0,\omega, -\omega)$, \markup{as they essentially become indistinguishable}. Using the diagrammatic approach, one can construct a decomposition into terms which will yield cancellations, and allow the observation of the convergence of the \nd~ harmonic and photovoltaic tensors. To obtain this Eqs.~\eqref{eq:U}-\eqref{eq:v1} are rewritten for the case $\bar{\omega} = 2\omega$, \markup{which reveals poles at} $\omega \pm \vare{n}{m}$ and $2\omega \pm \vare{n}{m}$. As we expect cancellations in the regime where the response is expected to vanish, we focus only on the most dominant, double photon $2\omega \to \vare{n}{m}$ pole. Since $2 \omega  \sim  \vare{n}{m}$, it is true that $\omega < \vare{n}{m}$, as well. For frequencies $\omega < E_g$, the former $2 \omega$ pole is the only process which may significantly participate, and yield resonant contributions. For simplicity, we only present results for $\sigma^{aa;c}$, as the most relevant contribution for linear-polarized light. The pole analysis is carried out in the following manner. The two-legged diagram in Eq.~\eqref{eq:U} contains a pole whenever $\bar{\omega} = -\vare{n}{m}$. This resonance is encountered Eq.~\eqref{eq:v1} whenever $l=m$, constraining the two-photon process to be essentially a two-band process, even in the velocity gauge. Elsewhere, the resonance condition gives that $\omega_{1,2} = \frac{\vare{n}{m}}{2}$, but since the overall occupation factor $f_{nm}$ prevents denominators with $\vare{n}{m}^{-1}$ from diverging, all contributions without the overall $\bar{\omega}^{-1}$ pole are subleading. Analyzing the resonances, we construct analogues of the injection, and counter-injection conductivities for $\bar{\omega} = 2\omega$. These read,
		\begin{align}
			& \sigma^{aa;c}_{2\omega,\textrm{inj}} = \frac{e^3}{\hbar^2} \sum_{nm} \int_\bk f_{nm} N_{nm}(2\omega) \biggl(\frac{\Re{r^a_{nm}r^c_{mn}}}{\varepsilon_{nm}} \Delta^c_{mn} \biggr) \label{eq:2ores1} && \\ 
			\notag &\sigma^{aa;c}_{2\omega,\textrm{c-inj}} = -\frac{e^3}{2\hbar^2 } \sum_{nm} \int_\bk  f_{nm} N_{nm}(2\omega) \biggl(\frac{ |r^a_{nm}|^2 \Delta^c_{mn}}{\vare{m}{n}} ~ +  ~\\
			&  \frac{ \Re{r^a_{nm}r^c_{mn}}\Delta^a_{mn}}{\vare{m}{n}}  + r^a_{nm}\lambda^{ac}_{mn} - r^a_{nm} r^c_{mn} \delta^b_{mn} + \frac{r^a_{nm}}{\vare{n}{m}}\tlo^{ac,1}_{mn}\biggl).
		\end{align}
		Note that $N_{nm}(2\omega)$ depends on the broadening induced by a two photon process, and hence reads here $N_{nm}(2\omega) = (2\omega-\vare{n}{m} + i\alpha \tau^{-1})^{-1}$. The Fermi surface contribution is obtained by considering the subleading parts referred to above. Analyzing Eq.~\ref{eq:w1} for the case $\omega_{1,2} = -\frac{\vare{n}{m}}{2}$, we find this term gives $ \frac{v^a_{nm} w^{ac}_{mn}}{2\vare{n}{m}} = 
		\frac{f_{n}}{2}[r^a, w^{ac}]_{nn}$. Other cancellations follow by using the identities derived in Sec. \ref{sec:zerofreq}.  
		Combining this with the off-resonant part of Eq.~\eqref{eq:U} gives the $2\omega$ Fermi surface contribution,
		\begin{align}
			\sigma_{\textrm{FS}}^{aa;c} 
			&= -\frac{e^3}{2 \hbar^2 \omega_1 \omega_2} \int_\bk f_n \left(\partial_a^2 \partial_c \varepsilon_n - \partial_c \tlo^{ac,1}_{nn}\right).
			\label{eq:2o2ores2}
		\end{align}
		In the above, we only kept terms which are total derivatives and hence are found on the Fermi surface. 
		This result should be compared with the FS conductivity of the photovoltaic response. The integrand of Eq.~\eqref{eq:2o2ores2} is identical to the one in Eq.~\eqref{eq:fs}, but the prefactor $(\omega_1\omega_2)^{-1}$ is different between the photovoltaic and \nd~harmonic cases. Using the proper lifetime insertion prescription, we demand that $\omega_1 \to \omega + \frac{i}{\tau}, \omega_2 = \omega_1$. Thus, only in the true $\omega \to 0$ are both terms identical and consistent with each other. The crossover, therefore, must occur when $\omega \sim \frac{1}{\tau}$, and this is precisely the scale at which the \nd~harmonic and photovoltaic response begin to overlap. The correction to the $\omega \to 0$ limit appears in the small frequency expansion for the photovoltaic case $\sigma_{\textrm{FS}}(\omega_1 =-\omega_2) \sim -\tau^2 +\omega^2 \tau^4$, while $\sigma_{\textrm{FS}}(\omega_1 =\omega_2) \sim -\tau^2+ 3\omega^2 \tau^4$. (We omitted the linear term in the expansion, as it vanishes for linear-polarized light). Thus, the modulation of the conductivity at small but finite frequency experiences a cross-over, where the \nd~harmonic response departs from the photovoltaic limit above the scale of $\sim \frac{1}{\tau}$. We point out that the FS contribution outlined here was recently explored in a study of the band-geometric properties of \nd~ harmonic response in Ref.~\cite{Bhalla2021}.
		Using the FS contribution we have shown the consistency of the \nd~harmonic tensors and photovoltaic elements of the conductivity, in the limit $\omega \to 0$. 
		%We find that the FS and injection-like response occur in the direction of the field $a$, as opposed to the photovoltaic injection effect which is in the direction of the current probe, $c$.   
		In the limit of finite frequency, we integrate Eqs.~\eqref{eq:2ores1}-\eqref{eq:2o2ores2} numerically using the model in App. \ref{app:model}. The conductivities -- both resonant and off-resonant -- are shown in \markup{Fig.~\ref{fig:fig3}}, in a frequency window of $2\omega \sim E_g$. 
		\begin{figure}[h]
			\centering
			\includegraphics[width=1.0\columnwidth]{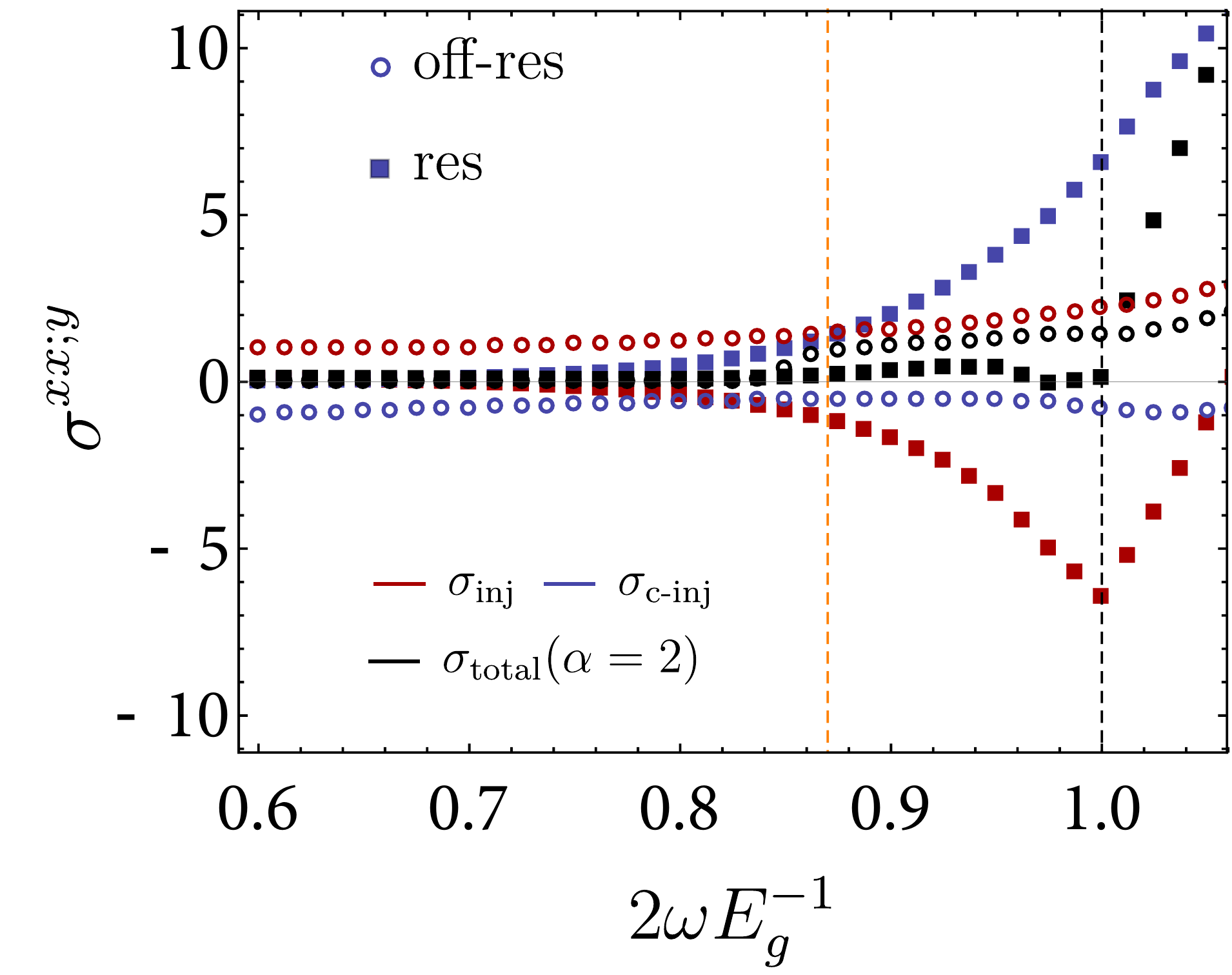}
			\caption{2nd harmonic nonlinear conductivity in the vicinity of $\bar{\omega} = E_g$. Squares denote resonant contributions, while empty circles are off-resonant terms. We find that in the vicinity of this resonance, injection, and counter-injection conductivities cancel each other, below the gap. Throughout, $\alpha=2$, and the chemical potential $\mu = 0$, making the Fermi surface contribution identically zero.} 
			\label{fig:fig3}
		\end{figure}
	}
	\section{Beyond semiclassics}
	\label{sec:beyondsemi}
	\markup{As an extension of the procedure in Sec.~\ref{sec:zerofreq}, we consider the continuation of the series in $\tau$}, for $\tau^n, n < 0$. The motivation for continuing the series stems from the fundamental difference between the limit $\omega \to 0$ in the diagrammatic formalism, versus the similar treatment in the Boltzmann approach. In the Boltzmann picture, the relaxation time $\tau$ refers to the inverse rate of the relaxation of the distribution function $f(\bm{r},\bm{k},t)$ to some equilibrium state, according to the relaxation time approximation. In the quantum perturbative formalism $(2\tau)^{-1}$ refers to the self-energy of the quasiparticles in the single-particle approximation, which is derived from multiband processes in the entire Brillouin zone. In order to allow a meaningful comparison between the two approaches, we assume for the self energy $[2\tau(\omega, \bm{k})]^{-1} \approx \textrm{const}$~\cite{Cheng2015,Holder2020}. This is similar in the Boltzmann approach, in which the collision integral is replaced by the relaxation time approximation, $\propto \frac{f_n-f^{0}_n}{\tau}$, where $\tau = \textrm{const}$. In the Boltzmann description, conductivities attain powers of $\tau$ through the expansion of the non-equilibirum distribution function $f(\bm{r},\bm{k},t)$ in the applied perturbation. In the diagrammatic case, they appear in the expansion of the dressed electron propagator, in the limit $\tau \to \infty$. The Boltzmann approach enforces that $\tau^n$ must have $n \ge 0$, as the power in $\tau$ is directly related to the order in the electric field $E^a$ which is responsible for the perturbation of the distribution function. Indeed, negative powers of $\tau$ are therefore impossible \cite{Mahan1990}. However, the diagrammatic formalism imposes no such restriction, as the origin of the $\tau$ dependence is entirely different, as explained above. After expanding the total conductivity in App. \ref{app:D} to order $\tau^{-1}$, with the identities of Sec. \ref{sec:zerofreq}, we arrive at the following conductivity,
	\begin{align}
		\sigma_{3}^{ab;c} = -\frac{e^3}{\hbar^2} \frac{1}{\tau} \sum_n \int_{\bk} f_n \left(\partial_a \tlo^{bc,-2}_{nn} + \partial_b \tlo^{ac,-2}_{nn} \right),
		\label{eq:taum}
	\end{align}
	where we already assumed $\alpha = 2$ for simplicity. In order to obtain this result, we also relied on additional identities,
	\begin{align}
		\notag 
		&\partial_c [\varepsilon^{-2} r^a, r^b]_{nn} = \frac{1}{2} [\varepsilon^{-2} r^a, \tlo^{cb,0}]_{nn}  -\frac{1}{2} [\varepsilon^{-2} r^b, \tlo^{ca,0}]_{nn} \\ 
		& +  [\varepsilon^{-2} r^a,\lambda^{cb}]_{nn} - [\varepsilon^{-2} r^b,\lambda^{ca}]_{nn} + [\varepsilon^{-3} r^a,\Delta^c r^b]_{nn} ,
	\end{align}
	and the Jacobi identity,
	\begin{align}
		[\varepsilon^{-2} r^a, \tlo^{bc,0}]_{nn} + [r^c, \tlo^{ab,-2}]_{nn} - [r^b, \tlo^{ac,-2}]_{nn} = 0.
	\end{align}
	Firstly, we observe that Eq.~\eqref{eq:taum} describes a transverse response, similar to that of the Berry curvature dipole (Eq.~\eqref{eq:tau1}). The properties of this term under crystal symmetries and time reversal are elaborated upon in Sec.~\ref{sec:discuss}. 
	\begin{figure}[h]
		\centering
		\includegraphics[width=1.0\columnwidth]{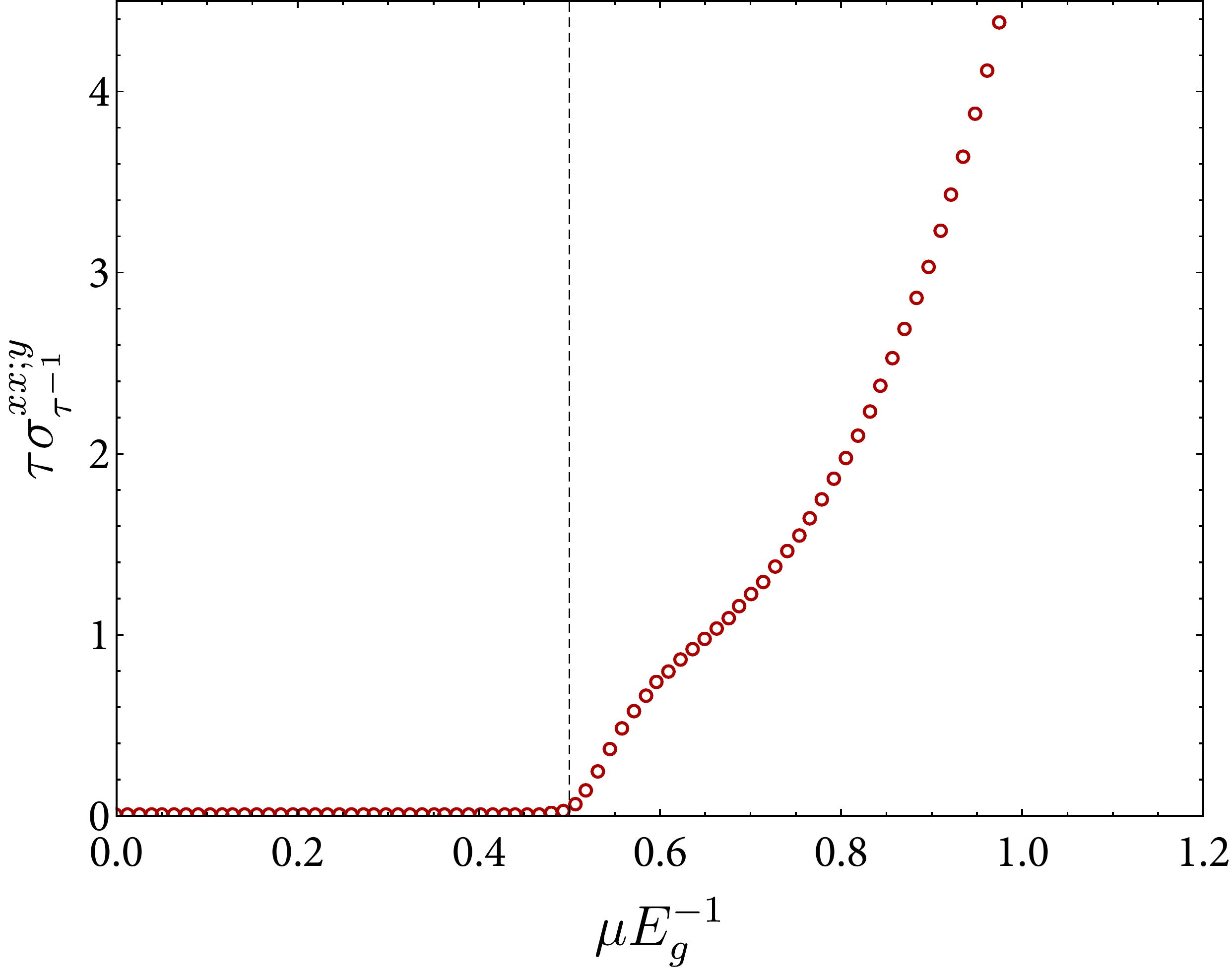}
		\caption{Transverse conductivity at order $\tau^{-1}$. Dashed lines indicate the onset of a band gap. The conductivity rises monotonically whenever $|\mu| > E_g$, and the result exhibits particle hole symmetry, $\sigma_{3}^{xx;y}(\mu) = \sigma_{3}^{xx;y}(-\mu)$. This is explained in App.~\ref{sec:particle_hole}.} 
		\label{fig:fig5}
	\end{figure}
	In Fig.~\ref{fig:fig5}, using the model of App.~\ref{app:model} we plot $\sigma^{xx;y}_{3}$ as a function of $\mu$. We observe, as expected from the form of Eq.~\eqref{eq:taum}, that only Fermi surface contributions enter the expression. Moreover, in the same chemical potential window as in Fig.~\ref{fig:fig4}, this term appears to monotonically increase, and obeys particle-hole symmetry, when $\mu \to -\mu$. We note that this term is naturally expected to vanish in the clean limit $\tau \to \infty$, and has therefore been overlooked in numerical computations of the full conductivity, Eq.~\eqref{eq:g2nd}.
	
	\section{Discussion}
	\label{sec:discuss}
	We now comment on some consequences for the phenomenology of \nd~order optical response.
	{\allowdisplaybreaks Our main results for the dc conductivity are Eqs.~\eqref{eq:tau2}-\eqref{eq:tau0}. These equations present the result of cancellations in a systematic expansion in $\tau$, according to the lifetime prescription of Sec.~\ref{sec:lifetimes}. 
		Eqs.~\eqref{eq:tau2}-\eqref{eq:tau0} depend on the lifetime ratio $\alpha$. 
		A ratio $\alpha \neq 1$ should be viewed as a description which distinguishes between multi-photon and single-photon processes. Previous reports \cite{Cheng2017, Passos2018,Mikhailov2016,Cheng2015} have all concluded that the optical response depends sensitively on $\alpha$, irrespective of the choice of gauging. In particular, the value $\alpha=2$ has been singled out in several works as the correct choice for removing unphysical features in resonances \cite{Passos2018,Cheng2015,Parker2019}. Inserting $\alpha=2$ into our expressions, the dc-conductivity at second order becomes particularly simple, with the first four orders in $\tau^{-1}$ being
		\begin{align}
			\sigma_{\tau^2}^{ab;c} &= - \frac{e^3}{2\hbar^2} \tau^2 \int_\bk \sum_n  f_n 
			\partial_a\partial_b\partial_c \varepsilon_n + (a \lra b) \label{eq:tau2_c}\\ \
			\sigma_{\tau^1}^{ab;c} &= - \frac{e^3}{\hbar^2} \tau \int_\bk \sum_n  f_n
			\partial_b\tlo^{ac,0}_{nn} +  (a \lra b) \label{eq:tau1_c} \\ 
			\sigma_{\tau^0}^{ab;c} &= - \frac{e^3}{\hbar^2} \int_\bk \sum_n  f_n
			\left(\partial_c\tlo^{ab,-1}_{nn} - \frac{1}{2}\partial_a \tlo^{bc,-1}_{nn} + \eta^{ab;c}_{nn}\right)  \label{eq:tau0_c} \notag \\ & \qquad  + (a \lra b) \\ \sigma_{\tau^{-1}}^{ab;c} &= -\frac{e^3}{\hbar^2 \tau} \int_\bk \sum_n f_n  \left(\partial_a \tlo^{bc,-2}_{nn}  + (a \lra b) \right) \label{eq:taum1_c} 
		\end{align}
	}
	We note that this result overlaps entirely with semiclassical calculations of the nonlinear conductivity for $\sigma_{\tau^2}^{ab;c}$, $\sigma_{\tau^1}^{ab;c}$ and the first term in $\sigma_{\tau^0}^{ab;c}$, which have been reported before in the literature~\cite{Sodemann2015,Gao2019,Watanbe2020}. At the same time, we find that the semiclassical expressions reported up to now have been incomplete, missing an important piece in $\sigma_{\tau^0}^{ab;c}$. As this additional contribution, in the form of $\eta^{ab;c}_{nn}$ is non-dissipative (scales with $\tau^{0}$) and Hall-like (vanishes when $a = b = c$), we explore it in a separate work \cite{Kaplan2022a}. We have also derived a new contribution $\sigma_{\tau^{-1}}^{ab;c}$ which is inaccessible in the semiclassical approach.
	The mathematical congruence which we demonstrate seems somewhat expected, but we emphasize that such a finding has nonetheless been considered improbable in the literature. In particular, it has been argued that the velocity gauge cannot be extended to the $\omega \to 0$ in any meaningful way \cite{Taghizadeh2017}. 
	As we described in detail, the dc-limit is indeed not without subtleties. We find that without enforcing the specific ratio $\alpha=2$ for the lifetimes, the continuation of the quantum perturbative descriptions to $\omega = 0$ will result in additional terms for the conductivity at all orders in $\tau^{-1}$. The issue of fixing $\alpha$ appropriately for single-particle models is further discussed in App.~\ref{app:alpha}.
	Nevertheless, we strongly believe that even the case $\alpha \neq 2$ has physical significance, as the conductivity then becomes sensitive to the interplay of intraband and interband quasiparticle lifetimes, and could thus serve as a probe of interaction physics~\cite{Kaplan2020}. 
	
	We briefly comment on the behavior of every order in $\tau$ under TRS. An inspection of Eqs.~\eqref{eq:tau2_c}-\eqref{eq:tau0_c} reveals that all terms of order $\tau^{2n}$ are odd under TRS, whereas the ones of size $\tau^{2n+1}$ are even. 
	This is is true also for the new term $\sigma_{\tau^{-1}}$, which is even under TRS, and therefore does not contain, when $\alpha = 2$, any in-gap response, as expected. This finding can be generalized to arbitrary order in the nonlinear response: for third order, for example, the real part of band-resolved objects is now even (as it is a quadruple product), while the imaginary part is odd under TRS.
	Therefore, the leading order in this case, of size $\tau^{3}$ is even under TRS, followed by the odd term at order $\tau^2$, as is known from studies of the 3\textsuperscript{rd} order conductivity in graphene~\cite{Mikhailov2014}. 
	
	Finally, we have demonstrated that cancellations are not only relevant for the dc-conductivity, but also play a role when discussing resonances at finite frequencies. 
	Starting from the well-established observation that for finite lifetimes, the optical response cannot vanish identically inside a single-particle gap, we derived the remainder terms from the photovoltaic and second harmonic resonances. 
	We found for both that they are composed of two counteracting pieces.
	In the case of the photovoltaic effect, these counteracting contributions are the leftovers from the well-known injection and shift currents.
	Furthermore, the photovoltaic effect is found to be sensitive to the ratio $\alpha$, whereas the \nd~ harmonic response (due to the absence of scale separation in $\tau$ of the pole structure) is not. 
	The energy scale for the onset of semiclassics -- where cancellations are expected to occur (Fig.~\ref{fig:fig2}(inset)) -- varies very little with $\alpha$. 
	We note that most previous works did not encounter these subtleties because TRS and other crystalline symmetries (such as in-place mirrors in spinless models) trivially remove the in-gap response. Additional rotational symmetries (such as $C_4$ on a simple cubic lattice) may also enforce cancellations where there otherwise would be a divergence. The question of which nonlinear couplings in a tight-binding formulation are needed in order to account for the emergence of higher-order nonlinear affects was recently explored in Ref.~\cite{Oiwa2022}.
	
	Finally, we note that the order of limits $\tau\to\infty$, $\omega\to 0$ leads to the semiclassical currents in Eqs.~(\ref{eq:fs},\ref{eq:2o2ores2}) for the photovoltaic and \nd~harmonic currents, which are identical in magnitude but carry opposite signs.
	In the dc-limit, the total conductivity is necessarily the sum $\sigma^{ab;c}(0) = \sigma^{ab;c}_{2\omega} +  \sigma^{ab;c}_{\textrm{photovoltaic}}=0$, thereby vanishing exactly.
	Obviously, the reverse order of limits $\omega\to 0$, $\tau\to\infty$ yields instead the non-zero Eqs.~(\ref{eq:tau2_c}-\ref{eq:taum1_c}).
	This non-commutativity of the limits is well-known, and is related to the crossover scale $\omega\tau\sim 1$ at which neither $\omega$ nor $\tau$ can be expanded independently. 
	Our results therefore resolve any conjectured discrepancy between semiclassical and quantum perturbative approaches~\cite{Ventura2017}.
	It is probably only due to the simplicity of the expressions for linear response that an incompatibilty of both approaches has been proposed, simply because in the latter case, both limits can be easily obtained from each other by the replacement $\omega\to\tau^{-1}$. 
	As demonstrated here in detail, such is no longer a good prescription beyond linear order, and in fact leads to incorrect expressions. 
	
	To show the power of our algebraic approach, we used the diagrammatic formalism to calculate the leading quantum contribution to the \nd-order conductivity, which scales as $\tau^{-1}$. As it is even under TRS, it should be observable in dirty nonmagnetic systems, provided there is a Fermi surface in the $\omega \to 0$ limit. By measuring a sample at different temperatures, the $\tau$ scaling should reveal a contribution which increases with disorder strength and temperature. A detection of such a term, at nonlinear order in the electric field, can be explained only through the procedure outlined in this work.
	
	We comment briefly on the prospects of measuring Eq.~\eqref{eq:taum1_c}. The main difficulty in detecting an object of order $\tau^{-1}$ are the other (superleading) parts of the conductivity. A recent proposal \cite{Ledwith2020,Xie2021} suggest that the chiral limit of twisted bilayer graphene (TBG) will exhibit an exactly flat Berry curvature. For integer fillings, is it also expected to be time-reversal symmetric. 
	Following a similar idea for linear response~\cite{Mitscherling2022}, we point out that for flat bands Eqs.~(\ref{eq:tau2_c}-\ref{eq:tau0_c}) vanish identically, leaving only Eq.~\eqref{eq:taum1_c}, provided at least a small Fermi surface is present. Other flatband materials could be suitable candidates, if the Berry curvature dipole contribution could be avoided~\cite{Varjas2022}. 
	
	Before closing, since we did not include in our analysis extrinsic effects \cite{Niu2019,Du2021}, we point out a possible route to subtract extrinsic contributions from the nonlinear current. To this end, let us assume that the band electrons scatter from isotropic impurities. In that case, the dc-conductivity will retain the cyclic permutation symmetry in spatial indices, i.~e. $\sigma^{ab;c} = \sigma^{c;ab} = \sigma^{b;ca}$. Therefore, the difference $\sigma^{ab;c}-\sigma^{c;ab}$ will eliminate the effect of isotropic scatterers, and leave only the quantum anomalous terms. These include $\sigma_{\tau^0}^{ab;c}$, as well as the newly derived $\sigma_{\tau^{-1}}^{ab;c}$.
	
	In summary, in this work we have outlined a consistent procedure through which the dc limit may be extracted from quantum perturbative calculations at finite frequency. This limit strongly depends on the ratio of lifetimes introduced in poles which appear in the response formula. At \nd~order, this is the specifically the single photon $\omega_{1,2} - \vare{n}{m}$ and double photon $\omega_{1} + \omega_2 - \varepsilon_{nm}$. We have shown that the ratio $\alpha = 2$ appropriately reproduces semiclassical formulae, while also predicting new terms and features. This result can be generalized to all higher orders in the electric field; for the $n-th$ order, there should appear 1, 2, \ldots, $n$ frequency poles, which would require the replacement with $\frac{i}{\tau}, \frac{2i}{\tau}, \frac{3i}{\tau}, \ldots$, for a direct comparison with semiclassical calculations. Furthermore, new terms may appear which go beyond semiclassics. 
	%At \nd~order, this is given by Eq.~\eqref{eq:taum1_c}. Beyond this, higher-order terms are expected. At any event, the terms with $\alpha \neq 2$ can be used to probe new physics, and the sensitive interplay of lifetimes and interactions, as was previously proposed in Ref. \cite{Kaplan2020}.
	
	\begin{acknowledgements}
		B.Y.\ acknowledges the financial support by the European Research Council (ERC Consolidator Grant No. 815869, ``NonlinearTopo'') and Israel Science Foundation (ISF No. 2932/21). 
		D. Kaplan appreciates support from the Weizmann Institute Sustainability and Energy Research Initiative. D. Kaplan wishes to thank Joel E. Moore for enlightening discussions.
	\end{acknowledgements}
	% \bibliography{lit}
	%\onecolumngrid
	% \newpage
	\appendix
	\section{Fixing $\alpha$}
	\label{app:alpha}
	We elaborate on the process of fixing $\alpha$ to a particular value, for a given model. We first note that in a fully interacting system, it should be possible to determine $\alpha$ simply by considering the imaginary part of the self-energy of the interacting Green's function, without assuming the relaxation time approximation, that is, without invoking $\langle \Sigma_n(\omega, \mathbf{k}) \rangle = \frac{i}{2\tau_n}$ \cite{Holder2020}. The problem of determining $\alpha$ arises in single-particle models where the self-energies are usually replaced by constants, which are put equal for all bands. Following Ref.~\cite{Kaplan2020}, we first note the impossibility of maintaining all relaxation times to be identical, for all bands. In the simplest case, we consider a minimal two band system in which we assign the valence band a quasiparticle lifetime of $\tau_v$ and the excited state a lifetime of $\tau_c$. Then, the \nd~order relaxation rate in the valence band is $\gamma = \frac{1}{2\tau_v} + \frac{1}{2\tau_v} = \frac{1}{\tau_v}$. For a \nd~order process involving an interband relaxation (which consists of an excitation and propagation in the valence band) we assign the rate $\Gamma = \frac{1}{2\tau_v} + \frac{1}{2\tau_c}$. In this minimal model, $\alpha = \frac{\gamma}{\Gamma}$, 
	\begin{align}
		\alpha = \frac{\frac{1}{\tau_v}}{\frac{1}{2\tau_v} + \frac{1}{2\tau_c}}.
	\end{align}
	Next, we apply the \emph{definition} of semiclassical transport. In the semiclassical limit, all relaxation times are assumed to tend to the clean limit, $\tau_v, \tau_c \to \infty$ but all quasiparticles bands are sharp (clean), and therefore it must hold simultaneously that $\frac{\tau_c}{\tau_v} \to \infty$. In this limit $\alpha \to 2$, and this formally gives our conclusion, presented in Sec.~\ref{sec:discuss} that only $\alpha = 2$ relates the true semiclassical limit. Note that the assumption $\tau_v = \tau_c \to \infty$ would instead set $\alpha = 1$ irrespective of taking the clean-limit $\tau \to \infty$, which would result in additional, \emph{non}-semiclassical terms contributing to Eqs.~\eqref{eq:tau2}-\eqref{eq:tau0}.
	Alternatively, it  is possible to fix $\alpha$ by considering the distance in frequency at which the semiclassical regime sets on (for finite frequency), such that frequency is below the fundamental gap. This is illustrated in the inset of Fig.~\ref{fig:fig2}.
	\section{Real and imaginary parts}
	\label{app:A}
	In order to determine which contributions enter the response, it is necessary to identify precisely whether they are real, or imaginary. The first immediate overall condition is that the conductivity $\sigma^{ab;c}$ is purely real (for linear polarized light). We start by focusing on the overall pole dependence seen in Eqs.~\eqref{eq:U}-~\eqref{eq:v1}. The global pole $(\omega_1 \omega_2)^{-1}$ behaves as,
	\begin{align}
		&\Re{\frac{1}{(\omega + i\tau^{-1})(-\omega + i\tau^{-1})}} = -\frac{\tau^2}{(\omega^2 \tau^2 +1)}, \\ 
		&\Im{\frac{1}{(\omega + i\tau^{-1})(-\omega + i\tau^{-1})}} = 0, \label{eq:app_photovol}
	\end{align}
	While for the \nd~ harmonic,
	\begin{align}
		&\Re{\frac{1}{(\omega + i\tau^{-1})(\omega + i\tau^{-1})}} = \tau^2 \frac{\omega^2 \tau^2 -1}{(1+\tau^2\omega^2)^2}, \\ 
		&\Im{\frac{1}{(\omega + i\tau^{-1})(\omega + i\tau^{-1})}} = - 2\tau^3 \omega \frac{1}{(1+\tau^2\omega^2)}
	\end{align}
	Importantly, the imaginary part of the \nd~harmonic vanishes identically at $\omega \to 0$, since by global symmetry constraints it cannot survive. We also observe that the contribution of the overall pole is identical for both the \nd~harmonic and photovoltaic cases, in the strict $\omega \to 0$ limit, and become indistinguishable in the vicinity of $\omega \tau \sim 1$. Next, one must find expressions for the denominators involving energy differences. This takes the form, in the photovoltaic limit,
	\begin{align}
		\notag  &\Re{\frac{1}{\left(\omega+i\tau^{-1}\right)\left(-\omega+i\tau^{-1}\right)\left(\omega+\varepsilon_{nm}+i\tau^{-1}\right)}} \\ 
		& =\frac{\tau ^4 \left(\varepsilon _{{nm}}+\omega \right)}{\left(\tau ^2 \omega ^2+1\right) \left(\tau ^2 \varepsilon _{{nm}} \left(\varepsilon _{{nm}}+2 \omega \right)+\tau ^2 \omega ^2+1\right)}.
	\end{align}
	After setting $\omega \to 0$, and expanding in powers of $\tau$, one has,
	\begin{align}
		\notag & \lim_{\omega \to 0} \frac{\tau ^4 \left(\varepsilon _{{nm}}+\omega \right)}{\left(\tau ^2 \omega ^2+1\right) \left(\tau ^2 \varepsilon _{{nm}} \left(\varepsilon _{{nm}}+2 \omega \right)+\tau ^2 \omega ^2+1\right)} = \\ &
		\frac{\tau^2}{\varepsilon_{nm}} - \frac{1}{\varepsilon_{nm}^3} + \mathcal{O}(\tau^{-2}). \label{eq:single_pole}
	\end{align}
	Particular care should be taken when additional pole appears in the denominator. While Eq.~\eqref{eq:single_pole} was thus expandable because a global Fermi-Dirac factor of $f_n - f_m$ prevents a zero frequency divergence, an additional unrestricted denominator does not have this property. One encounters the combination,
	\begin{align}
		\notag &\mathrm{Re}\Bigl[
		\left(\omega+i\tau^{-1}\right)^{-1}
		\left(\omega-i\tau^{-1}\right)^{-1}
		\notag\\
		&\quad\left(\omega+\varepsilon_{nm}i\tau^{-1}\right)^{-1}
		\left(0+\varepsilon_{nl}+i\alpha\tau^{-1}\right)
		\Bigr]\\ 
		& \overset{\omega \to 0}{=} \left\lbrace
		\begin{matrix}
			\frac{1}{\alpha  \varepsilon _{{nm}}^4}-\frac{\tau ^2}{\alpha  \varepsilon _{{nm}}^2} + \mathcal{O}(\tau^{-2}), & n = l \\ 
			\frac{\tau ^2}{\varepsilon _{{nl}} \varepsilon _{{nm}}}-\frac{\alpha  \varepsilon _{{nl}} \varepsilon _{{nm}}+\varepsilon _{{nl}}^2+\alpha ^2 \varepsilon _{{nm}}^2}{\varepsilon _{{nl}}^3 \varepsilon _{{nm}}^3} + \mathcal{O}(\tau^{-2}), & n \neq l.
		\end{matrix}
		\right.
		\label{eq:doublepole}
	\end{align}
	Additionally, Eq.~\eqref{eq:U} explicitly contains a pole at $\omega_{1}+\omega_{2}$ which requires separate treatment in the photovoltaic case. This takes the form,
	\begin{align}
		\notag
		&\lim_{\omega \to 0} \frac{1}{\left(\omega -\frac{i}{\tau }\right) \left(\omega +\frac{i}{\tau }\right) \left(\varepsilon _{{nm}}+\frac{i \alpha }{\tau }\right)} = \\ &
		\frac{\tau ^2}{\varepsilon _{{nm}}}-\frac{\alpha ^2}{\varepsilon _{{nm}}^3} + \mathcal{O}(\tau^{-2}).
	\end{align}
	An overall real conductivity can be attained by considering the imaginary part of the Green's functions coupled with the imaginary part of the matrix elements of the response. The contact terms, such as the one in $\mathcal{U}^{ab;c}$ appearing without any propagators vanish according to Eq.~\eqref{eq:app_photovol}. However, once a propagator is present, the following combination is relevant,
	\begin{align}
		\lim_{\omega \to 0}\Im{\frac{1}{\left(\omega -\frac{i}{\tau }\right) \left(\omega +\frac{i}{\tau }\right) \left(\varepsilon _{{nm}}+\frac{i \alpha }{\tau }\right)}} = \frac{\alpha \tau}{\varepsilon_{nm}^2} + \mathcal{O}(\tau^{-1}).
	\end{align}
	The requirement of the imaginary restricts whether the matrix elements contribute; this is strongly affected by spatial symmetries. Combinations such as $w^{aa}_{nm} v^a_{mn}$ will not produce an imaginary part, and hence will vanish at this order in $\mathcal{O}(\tau)$. This explains how only odd terms under the exchange $(a \lra c)$ (such as the Berry curvature dipole) appear at order $\tau$.
	Finally, additional contributions can be considered for $\mathcal{V}^{ab;c}$. For the photovoltaic case, one will have a leading order divergence which is $\sim \tau^3$ whose matrix elements are $v^a_{nm} v^b_{mn} v^c_{nn}$, which is odd under complex conjugation. After symmetrization with respect to $(a \lra b)$ and taking the real part (which is the only surviving DC component) in the $\omega \to 0$ limit, these terms vanish altogether. We are left with the case,
	\begin{align}
		\notag &\mathrm{Im}
		\biggl[\frac{1}{\left(\omega+i\tau^{-1}\right)\left(\omega-i\tau^{-1}\right)\left(\omega+\varepsilon_{nm}i\tau^{-1}\right)\left(\varepsilon_{nl}+i\alpha\tau^{-1}\right)}\biggr]
		\\ & = \frac{\alpha \tau}{\varepsilon_{nm}^2\varepsilon_{nl}^2}\left(\varepsilon_{nm}+\varepsilon_{nl}\right)  + \mathcal{O}(\tau^{-1})
	\end{align}
	The sum of $\varepsilon_{nm} + \varepsilon_{nl}$ can be used to construct transformation properties of $\tlo^{ab,1}$, shown in Sec.~\ref{sec:zerofreq}. 
	The elucidation of the real and imaginary parts of the Green's function are related to the properties of the underlying quasiparticles in the theory, in the usual way~\cite{Mahan1990}. 
	\section{Model Hamiltonian}
	\label{app:model}
\begin{figure}[h]
		\centering
		\includegraphics[width=0.8\columnwidth]{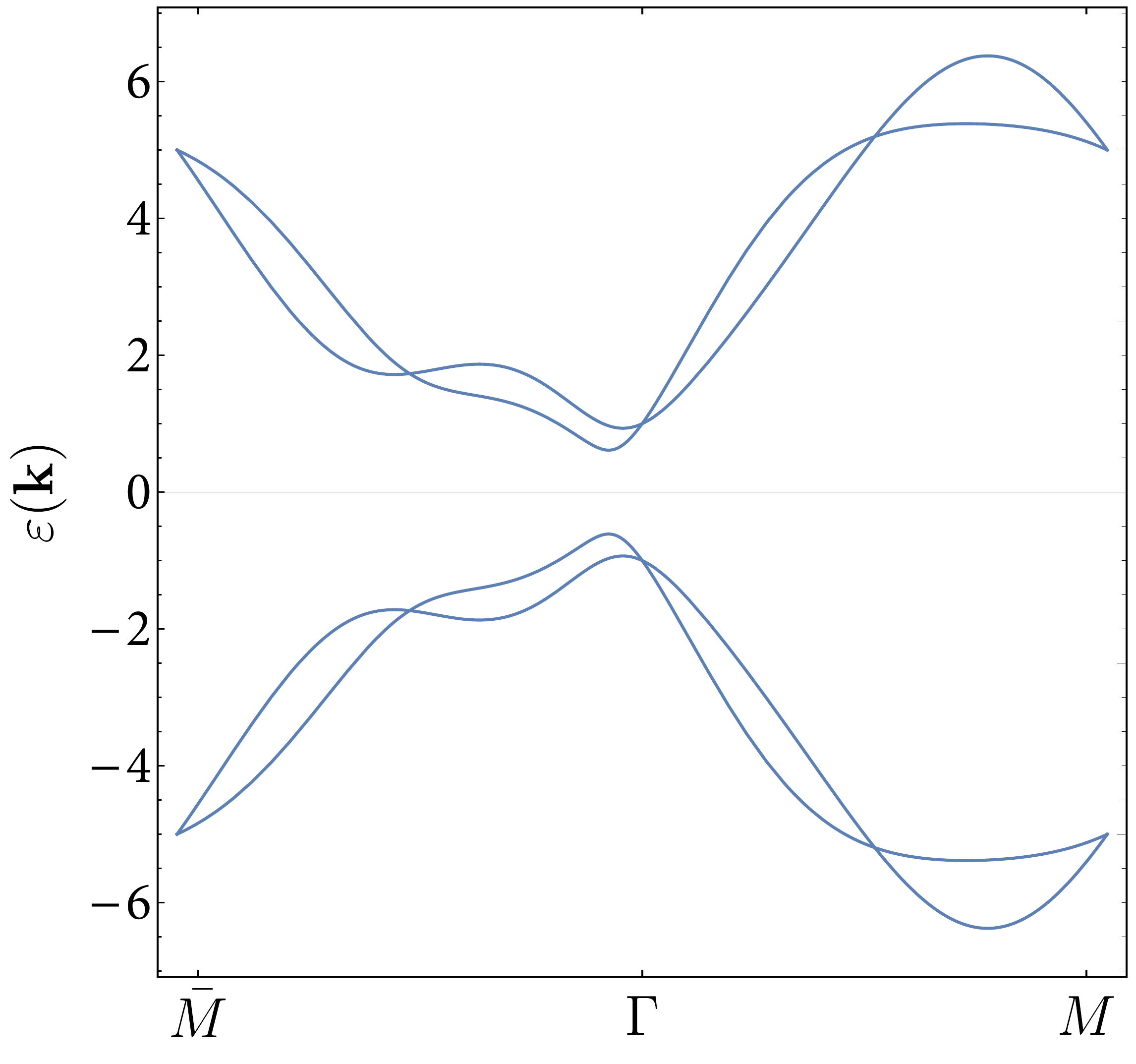}
		\caption{Band structure of the model used in this work, with $A = 1, B = 1, C=1$. $M = 3.5$. The chemical potential is fixed at $\mu = 0$. For these parameters, the gap $E_g \approx 0.5$, with direct transitions occurring along the $\bar{M} - \Gamma$ line.}
		\label{fig:band_structure}
	\end{figure}

	In the main text, we provide numerical results for the conductivites in Eqs.~\eqref{eq:tau2}-\eqref{eq:tau0} using the following tight-binding Hamiltonian $H$, 
	\begin{flalign}
		\frac{H}{t} &= (M-\cos(k_x)-\cos(k_y)) \sigma_z\tau_z  + \sin(k_x) \tau_x \sigma_z 
		\notag \\ 
		&\quad 
		+ \sin(k_y) \tau_y \sigma_z  
		+ \frac{C}{2}\left(\sin(k_x)+\sin(k_y)\right)\sigma_z \tau_z
		\notag\\ 
		&\quad 
		+  \frac{B}{2} \left(\sin(k_x+k_y) + \sin(k_x-k_y)\right)\sigma_y \tau_y 
		\notag\\ 
		&\quad 
		+ \frac{A}{2} \left(\cos(k_x +k_y)
		+\cos(k_x -k_y)\right)\sigma_x \tau_z .
		\label{eq:modelham}
	\end{flalign}
	This model describes an insulator that breaking inversion. Without loss of generality, we choose $A = 1, B = -1, C = 1$. The model breaks both inversion and time-reversal symmetries, represented by $\mathcal{P} = i\tau_z (k \to -k)$, and $\mathcal{T} = i\sigma_y (k \to -k) \mathcal{C}$, where $\mathcal{C}$ is complex conjugation. The resulting band structure along the $\Gamma-M$ line is shown in Fig.~\ref{fig:band_structure}.
	\section{Zero frequency conductivity}
	\label{app:D}
	For clarity, we present the full expression for the zero frequency conductivity, as it appears, for general $\alpha$.  
	\begin{widetext}
		\begin{align}
			\notag \sigma^{ab;c} &= -\frac{e^3}{\hbar^2} \int_\mathbf{k} \biggl[\sum_n f_n  \tau^2 \frac{u_{nn}^{ab;c}}{2} + \frac{\tau^2}{2}\sum_{nm} f_{nm} \frac{w^{ab}_{nm}v^c_{mn}}{\varepsilon_{nm} + i\alpha \tau^{-1}} 
			\tau^2\sum_{nm} f_{nm}\frac{v^{a}_{nm}w^{bc}_{mn}}{\varepsilon_{nm} + i\tau^{-1}} + \\ &  \tau^2\sum_{nml} f_{nm} \frac{v^a_{nm} v^b_{mn} v^c_{ln}}{\left(\varepsilon_{nm}+ i\tau^{-1}\right)\left(\varepsilon_{nl} + i\alpha \tau^{-1}\right)} + \tau^2\sum_{nml} f_{nm} \frac{v^c_{nl} v^b_{lm} v^a_{mn}}{\left(\varepsilon_{mn}+ i\tau^{-1}\right)\left(\varepsilon_{ln} + i\alpha \tau^{-1}\right)}  \biggr].
		\end{align}
	\end{widetext}
	To arrive at the $\omega \to 0$ response formulae (Sec.~\ref{sec:zerofreq}), an expansion in $\tau \to \infty$ is needed. This is carried out directly through the identities of App.\ref{app:A}, given that the overall conductivity must be real. Components not restricted by the Fermi factor difference $f_{nm}$ must be expanded separately, covering the equal -- and unequal -- band index cases.
	\allowdisplaybreaks
	\section{Linear response}
	\label{app:lin}
	We demonstrate how the geometrical identities we have derived allow to directly reconstruct the standard form of linear response. At linear order only two diagrams contribute~\cite{Parker2019,Mahan1990}. The conductivity at frequency $\omega$ therefore reads,
	\begin{align}
		\sigma^{\alpha \beta} = \frac{ie^2}{\hbar} \int_\mathbf{k} \left[\sum_{n}f_{n} \frac{w^{\alpha \beta}_{nn}}{\omega} + \sum_{nm} f_{nm} \frac{v^\alpha_{nm} v^\beta_{mn}}{\omega(\omega - \varepsilon_{nm})}\right].
	\end{align}
	Linear response involves a single frequency. Hence, applying the prescription Eq.~\eqref{eq:prescrip} of Sec.~\ref{sec:lifetimes}, $\omega \to \omega + \frac{i}{\tau}$. Taking the limit $\omega \to 0$,
	\begin{align}
		\sigma^{\alpha \beta} = \frac{e^2}{\hbar} \int_\mathbf{k} \left[\sum_{n}f_{n} \tau w^{\alpha \beta}_{nn} + \sum_{nm} f_{nm} \frac{\tau v^\alpha_{nm} v^\beta_{mn}}{(-\varepsilon_{nm} +i/\tau)}\right].
	\end{align}
	The global Fermi factor $f_{nm}$ allows for a simple expansion in powers of $\tau, \tau \to \infty$.
	\begin{align}
		\frac{\tau}{(-\varepsilon_{nm} +i/\tau)} = -\frac{\tau}{\varepsilon_{nm}} - \frac{i}{\varepsilon_{nm}^2} + \frac{1}{\varepsilon_{nm}^3 \tau} + \mathcal{O}(\tau^{-2}).
		\label{eq:linear2}
	\end{align}
	Using Eq.~\eqref{eq:ww}, $w^{\alpha \beta}_{nn} = \partial_\alpha \partial_\beta \varepsilon_{n} - \tlo^{\alpha \beta,1}_{nn}$. The leading order of the 2nd term in Eq.~\eqref{eq:linear2} is $-f_{nm} \tau v^\alpha_{nm} v^\beta_{nm} \varepsilon_{nm}^{-1} = \tau f_{nm}\varepsilon_{nm}r^\alpha_{nm} r^\beta_{mn} = \tau f_{n} \tlo^{\alpha \beta, 1}_{nn}$ after interchanging $(m \lra n)$, and expanding $f_{nm}$. This latter element identically cancels the one derived from $w^{\alpha \beta}_{nn}$. The next order term is,
	\begin{align}
		-\sum_{nm} f_{nm}\frac{i}{\varepsilon_{nm}^2}v^\alpha_{nm} v^\beta_{mn} = -\sum_n f_n \Omega_{nn}^{\alpha \beta},
	\end{align}
	which is $\tau^{0}$ and survives in an insulator, when $\int_\mathbf{k} \Omega^{ab}_{nn} \neq 0$. Since $\tau$ is finite, $\tau^{-1}$ (and higher negative powers) appear,
	\begin{align}
		\sum_{nm} f_{nm} \frac{v^\alpha_{nm} v^\beta_{mn}}{\varepsilon_{nm}^3 \tau} = \frac{1}{\tau} \sum_n f_n \left[\frac{r^\alpha}{\varepsilon}, r^\beta \right]_{nn}.
	\end{align}
	This contribution is gauge invariant and matches that derived recently by adjusting the conductivity with broadened spectral functions \cite{Mitscherling2020,Mitscherling2022}.
	\section{Total conductivity, lifetimes, and limits}
	The inclusion of finite lifetimes allows for the direct computation of the $\omega \to 0$ limit, without any explicit reliance on sum rules as in Refs. \cite{Aversa1994,Aversa1995,Sipe2000,Nastos2006}. Cancellations are instead regulated by the parameter $\alpha$ introduced in the main text. The ambiguity in the $\omega \to 0$ limit can be illustrated by comparing several procedures previously adopted to treat the $\omega \to 0$ case. The approach adopted in Sec.~\ref{sec:lifetimes} allows for probing 3 limits of interest. First, we keep $\tau$ finite (adopting some numerical value) but allowing $\alpha$ to deviate from the semiclassical value of $\alpha = 2$. 
	\begin{figure}[h]
		\centering
		\includegraphics[width=0.9\columnwidth]{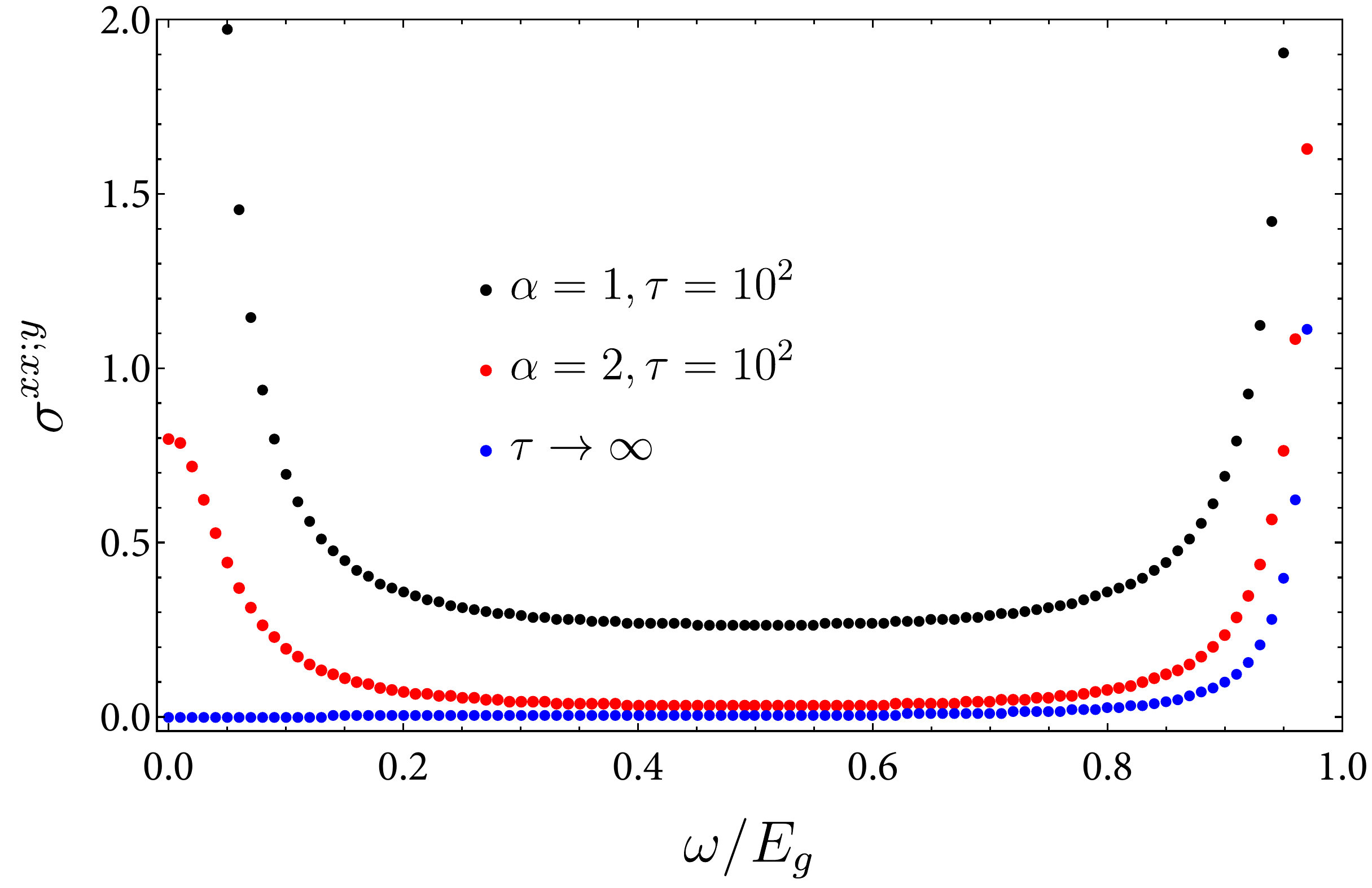}
		\caption{Hall conductivity as a function of frequency in the $\omega \to 0$ limit. Black: $\alpha = 1$ (setting all lifetimes equal). As a result, the transparent region increases (relative to the case of $\alpha = 2$. A seemingly physical enhancement not showing the characteristic Drude peak appears in the limit $\omega \to 0$. Red: $\alpha = 2$. In this case, the transparent region is more accurately represented (suppressed), and the $\omega \to 0$ shows the Drude peak features, in agreement with Ref. \cite{Michishita2021}. Blue: $\tau \to \infty$, where only resonant terms are kept. This limit, taken in Refs.~\cite{Sipe2000,Nastos2006} does not capture any semiclassical contributions. The total conductivity is calculated within the model of Eq.~\eqref{eq:modelham}, with the parameters $A=0.5, B=0.2, C=0, M=3.5$. The Fermi level is set to $\mu = -1.0$, and a finite Fermi surface exists.}
		\label{fig:regimes}
	\end{figure}
	In Fig.~\ref{fig:regimes}, we plot the full conductivity for frequencies below the resonance $\omega \sim E_g$, for a fixed value of $\tau$. We find that the relevant semiclassical features, such a suppressed transparency region for $\omega \gg \tau^{-1})$, and a pronounced Drude peak appear correctly, only when $\alpha = 2$. When $\alpha = 1$, for example, the Hall conductivity acquires stronger features, and is significantly larger, than in the $\alpha = 2$ case. Furthermore, the transparent region signal is now larger. While the approach to resonance appears similar, here too the onset of semiclassical behavior is modified, as discussed in Sec.~\ref{sec:bulkphotovoltaic}. Finally, one may consider the limit of $\tau \to \infty$, \emph{before} taking the $\omega \to 0$ limit. This is represented by the blue dots in Fig.~\ref{fig:regimes}. In this case, non-ambiguous results may be obtained only for the resonant components of the response, as shown in Ref. \cite{Sipe2000}. The result is that the response below the gap becomes zero, and all semiclassical effects vanish. 
	\section{Particle-hole symmetry and non-linear conductivities}
	\label{sec:particle_hole}
	In many tight-binding models, an apparent symmetry is the existence of reflection about the $\varepsilon(\mathbf{k}) = 0$ axis \cite{Bernevig2013}, such that valence and conduction bands are connected. Typically, this symmetry is represented by $\mathcal{C} = U \mathcal{K}$, where $\mathcal{K}$ is complex conjugation and $U$ is a model-specific unitary. The main property of this symmetry is that it is $k$-local, and is therefore relevant to all properties that are defined locally in k-space such as the conductivities of Eqs.~\eqref{eq:tau2}-\eqref{eq:tau0}. In the simple tight-binding model we used in Eq.~\eqref{eq:modelham}, this symmetry is represented by $\mathcal{C} = \tau_y \sigma_z \mathcal{K}$ and by its nature anti-commutes with the Hamiltonian. That is, $\lbrace H, \mathcal{C}\rbrace$ = 0, or $\mathcal{C} H(\mathbf{k}) \mathcal{C} = -H(\mathbf{k})$. Consider two states connected by this symmetry: $\left|n ' \mathbf{k}\right\rangle = \mathcal{C}\left|n  \mathbf{k}\right\rangle$. By definition, the eigenvalues of the two states are antipodean, i.e., $\varepsilon_{n}(\mathbf{k}) = - \varepsilon_{n'}(\mathbf{k})$. Any quantity which is a product of operators and purely real, however, is unchanged. For example $i[r^a, r^b]_{nn} \to \mathcal{C}i[r^a, r^b]_{nn} \mathcal{C} = i[r^a, r^b]_{nn}$ since this quantity is invariant under complex conjugation, but also since the product of operators is unchanged with the insertion of a unitary transformation $r^a_{nm} r^b_{mn} = [r^a U^\dagger U r^b]_{nn}$. Furthermore, the momentum derivative operator $\partial_a$ is invariant under $\mathcal{C}$. We can then formulate the following rules for the transformation of $\mu \to - \mu$, which is essentially the exchange of valence and conduction bands under $\mathcal{C}$,
	\begin{align}
		\sigma_{\tau^2}^{ab;c} (\mu) = - \sigma_{\tau^2}^{ab;c} (-\mu),\\
		\sigma_{\tau^{1}}^{ab;c} (\mu) = \sigma_{\tau^{1}}^{ab;c} (-\mu),\\
		\sigma_{\tau^0}^{ab;c} (\mu) = - \sigma_{\tau^{0}}^{ab;c} (-\mu),\\
		\sigma_{\tau^{-1}}^{ab;c} (\mu) = \sigma_{\tau^{-1}}^{ab;c} (-\mu).
	\end{align}
	This follows because Eqs.~\eqref{eq:tau2}-\eqref{eq:tau0} and Eq.~\eqref{eq:taum}, are separated by powers of the dispersion (and its derivatives). Terms of order $\tau^{2n}$ contain an odd product of dispersion and velocity elements, so they naturally gain a minus sign under $\mathcal{C}$, and are therefore odd under $\mu \to -\mu$, while terms of the form $\tau^{2n+1}$ contain an even number of dispersion elements and are therefore even under $\mu \to -\mu$. 
	\bibliography{main.bbl}

\begin{thebibliography}{61}
\bibitem [{b_K()}]{b_Kaplan2021}%
  \BibitemOpen
  \href@noop {} {\bibinfo {title} {{In velocity gauge, operators can only
  contain one power of $\varepsilon$. However, external powers of
  $\varepsilon$, such as those coming from resonance, may decrease (or
  increase) the power of $\varepsilon$ inside the commutator.}}}\BibitemShut
  {Stop}%
\end{thebibliography}
	
\end{document}